\documentclass[aps,preprint]{revtex4}%
\usepackage{amsfonts}
\usepackage{amsmath}
\usepackage{amssymb}
\usepackage{graphicx}%
\providecommand{\U}[1]{\protect\rule{.1in}{.1in}}

\begin{document}
\preprint{ }
\title[Short title for running header]{ ENTANGLEMENT AND GENERALIZED BERRY GEOMETRICAL PHASES IN QUANTUM GRAVITY}

\author{Diego J. CIRILO-LOMBARDO}
\affiliation{M. V. Keldysh Institute of the Russian Academy of Sciences, Federal Research
Center-Institute of Applied Mathematics, Miusskaya sq. 4, 125047 Moscow,
Russian Federation}
\affiliation{CONICET-Universidad de Buenos Aires, Departamento de Fisica, Instituto de
Fisica Interdisciplinaria y Aplicada (INFINA), Buenos Aires, Argentina.}
\author{Norma G. SANCHEZ}
\affiliation{International School of Astrophysics Daniel Chalonge - Hector de Vega, CNRS,
INSU-Institut National des Sciences de l'Univers, Sorbonne University  75014
Paris, France.\\
Norma.Sanchez@obspm.fr \\
https://chalonge-devega.fr/sanchez
}
\date{ \today}
\pacs{PACS number}

\begin{abstract}
A new formalism is introduced that makes it possible to elucidate the physical and geometric content of quantum spacetime. It 
is based in the Minimum Group Representation Principle (MGRP). Within this framework new results for entanglement and geometrical/topological phases are found and implemented in cosmological and black hole space-times. 
Our main results here are: 
{\bf(i)} We find the Berry phases for inflation and for the cosmological perturbations and express them in terms of the observables, as the spectral scalar and tensor indices, $n_S$ an $n_T$, and the tensor to scalar ratio $r$. The Berry phase for de Sitter inflation is imaginary with the sign describing the exponential acceleration.
{\bf(ii)} The pure entangled states in the  minimum group (metaplectic) $Mp(n)$ representation for quantum de Sitter space-time and black holes are found. {\bf(iii)} For entanglement, the relation between 
the Schmidt type representation and the 
physical states of the $Mp(n)$ group 
is found: This is a
{\it new non-diagonal} coherent state 
representation complementary to the known 
Sudarshan diagonal one. {\bf(iv)} Mean value generators of $Mp(2)$ are
related to the adiabatic invariant and  
topological charge of the spacetime, 
(matrix element of the 
transition $-\infty < t < \infty$). {\bf(v)} The basic {\it even} and {\it odd} $n$-sectors of the Hilbert space are intrinsic to the quantum spacetime and its discrete levels (in particular continuum for $n \rightarrow \infty$),  they  do not require any extrinsic generation 
process as the standard Schrodinger cat 
states, and are {\it entangled}. {\bf(vi)}  The gravity or cosmological domains on one side and another of the Planck scale are  {\it entangled}. Examples: The quantum primordial trans-Planckian de Sitter vacuum and the classical late de Sitter vacuum today; the central quantum gravity reqion and the external classical gravity region of black holes. The classical and quantum dual gravity regions of the space-time are entangled. {\bf(vii)} The general classical-quantum gravity duality is associated to the Metaplectic $Mp(n)$ group symmetry which provides the complete full covering of the phase space and of the quantum space-time mapped from it. 

\end{abstract}
\volumeyear{year}
\volumenumber{number}
\issuenumber{number}
\eid{identifier}
\date{\today}




\startpage{1}
\endpage{}
\maketitle
\tableofcontents

\section{Introduction and Results}

In recent years a deep interest has been manifested not only in the search for a consistent theory of quantum gravity but also for a fundamental description of the dynamics in the quantum domain. 

\medskip

Recently, in Refs. \cite{cirilo-sanchez}, \cite{universe} we have carried out a construction of the spacetime as a generalized coherent state of the $Mp(n)$ group (complying with the principle of the minimum group representation). 
We have shown that physical states are mappings of the group generators in a vector representation, and that these are expressed in the so called Sudarshan's
diagonal-representation and in a \textit{new non diagonal} one that leads, as an important consequence, the
\textit{physical states} with spin content 
 $(\,1/2,\;1,\;3/2,\;2\,)$. 

We have implemented this framework both in black holes and de Sitter space-time, showing explicitly this quantum fundamental dynamics. This approach goes well in the direction of Refs \cite{NSPRD2021}, \cite{NSPRD2023} for quantum space-time and its discrete levels, explicit examples and fundamental quantum principles: the classical-quantum duality of Nature including gravity (classical-quantum gravity duality), Refs. \cite{Sanchez2019a,Sanchez2019b,
Sanchez2019c}.

\medskip

Now, in this paper, let's take another step in this novel and consistent conceptual description of the universe considering entanglement mechanisms and their relationship with geometrical and topological phases (namely, generalized Berry phases). The groups involved here are non compact.

\medskip

Our formalism is fundamental to advance towards a true information theory of quantum gravity given that the entire theory is self-consistent and has important points in favor: Such as that the basic states that respond to the principle of least group representation satisfy a {\it new equation of positive energy} conceptually similar to those proposed by Majorana \cite{majorana1, majorana2} and Dirac \cite{dirac} time ago and that were very recently discussed by Bogomolny in Ref. \cite{bogomolni2022} related to the possible description of dark matter .
Precisely, our {\it generalized coherent states} here
generate a map that relates the metric $g_{ab}$, solution of the positive energy wave equation, 
to the specific subspace of the full Hilbert space where these coherent
states live.

\medskip

Taking advantage of our formalism, we find {\bf in this paper} the Berry phases for inflation and for the cosmological perturbations and express them in terms of the observables, as the spectral scalar and tensor indices, $n_S$ an $n_T$, and the tensor to scalar ratio $r$. The Berry phase for de Sitter inflation is imaginary with the sign describing the exponential acceleration.
For entanglement, {\bf we find here} the pure entangled states for de Sitter space-time as well as for black holes . 

\medskip

The complete covering of the Hilbert space with the complete covering of the quantum space-time is realized by the Metaplectic group symmetry which equivalently provides the complete global CPT symmetric states .  

We show here that:

The Classical-Quantum Duality of the space-time is also realized in the $ Mp(n)$ symmetry, because of the complete covering, the global complete space-time (and full phase space completion) is needed to make the classical-quantum space-time duality manifest (and the phase space mapped from it) .

\begin{itemize}

\item {It is worth to mention that similar discrete levels can be obtained from the 
 global (complete) classical - quantum  duality including gravity \cite{Sanchez2019a},\cite{Sanchez2019b},  \cite{Sanchez2019c}, \cite{NSPRD2021}, namely classical-quantum gravity duality  :   The two {\it even} and {\it odd}  (local) carts or sectors here and their global $(\pm)$ sum of states, reflect a relation between the $ Mp(n)$ symmetry and the classical-quantum gravity duality.} 

\item {The two $\sqrt{\,(2n+1)}$ and  $\sqrt{\,2n}$, {\it even} and {\it odd} sets separately are local coverings and they are {\it entangled}. The symmetric or antisymmetric sum of these states are global covering states, and they are necessary to cover completely the {\it whole} manifold.}  

\item {Moreover, the corresponding $(\,\pm\,)$ global states are complete, CPT symmetric and unitary, the levels $n = 0, 1, 2, ....,$ cover the whole Hilbert space $\mathcal{H}\; =\;\mathcal{H}_{\,(\, + \,)} \;\; \oplus\;\;\mathcal{H}_{\,(\,- \,)}$ and all the space-time regimes. In the Metaplectic group representation $ Mp (n) $, this  corresponds to the state sectors $\mathcal{H}_{\,(\, 1/4\,)}$ (even) and $\mathcal{H}_{\,(\, 3/4\,)}$} (odd), and as we find here these states {\it are entangled.}

\item {The total $n$ states range over \textit{all} scales from
the lowest excited levels to the highest excited ones covering the two dual branches $(+)$ and $(-)$ or Hilbert space sectors and corresponding space-time coverings.
The two $(+)$ and $(-)$ dual sectors are entangled.}

\end{itemize}

The consequences of these results are interesting because the classical-quantum gravity duality allows that signals or states in the quantum gravity (trans-Planckian) domaine, or semiquantum gravity (inflationary) domaine, do appear as low energy effects in the semiclassical/classical universe today. 

\medskip

From the results of this paper we see that the gravity domains from one side and the other of the Planck scale {\it are entangled }, for instance the black hole  quantum interior and the classical exterior gravity domains.
Similarly, for the cosmological domains: The quantum trans-Planckian primordial vacuum (constant curvature de Sitter phase, followed by a quasi-de Sitter (inflationary) phase), and the late classical dual de Sitter Universe today are {\it entangled}.

\medskip

We have also performed an interpretation of the quadratic quantum inflationary fluctuations in terms of the well-developed theory of time-dependent oscillators, their coherent, squeezed and cat states. Till now it remains unclear what are the consequences of such states for the cosmological evolution and the detection of CMB fluctuations, but hopefully such results will emerge from these investigations.

\medskip

Other new entanglement results of this paper are: 

\begin{itemize}
\item {The precise relation between the Schmidt type representation in the density matrix context and the physical state fulfilling the Minimal Group Representation Principle (MGRP), that is bilinear in the basic states of the $Mp(n)$ group, is found.} 

\item{ The mapping for the physical state refers to a {\it new non-diagonal} coherent state representation complementary to that of the known Sudarshan diagonal representation.} 
 
\item{The basic states in the Minimal Group Representation  sense: $|\,1/4>$ and $|\,3/4>$ (belonging to the even and odd sectors of the Hilbert space respectively) are intrinsic fundamental part of the very structure of the space-time itself and do not require an additional extrinsic generation process as in the standard Schrodinger cat states and their entanglement.} 
\end{itemize}

This paper is organized as follows: 

\medskip

In Section (II) we describe the Berry geometrical phases for non compact groups and the coherent state quantum evolution. In Section (III) we apply this framework to find the Berry phases for de Sitter inflation and  the inflationary (tensor and scalar) perturbations.  
Section (IV) is devoted to entanglement, its density matrix description and our generalized coherent states which allow to map the space-time metric into the Hilbert space. Section (V) deals with the $ Mp(n)$ coherent states describing the quantum evolution of the spacetime, the associated adiabatic invariants and the topological structure. In Sections (VI) and (VII) we find the entanglement with semi-coherent states and generalized Schrodinger cat states in the quantum space-time structure. In Section (VIII) we apply these results to find the entanglement in de Sitter and black hole space-times. Section (IX) is devoted to Remarks and Conclusions. 

\section{Geometrical Phases and Noncompact Groups}

\subsection{SU(1,1) coherent states and the Berry phase}%

Our starting point is the coset coherent state definition according to Perelomov-Klauder \cite{Kla1}, \cite{Kla}, \cite{pere} via the action of a displacement operator $D (\xi)$ belonging to the coset and generally unitary, as follows:
\[
\left\vert \;\psi \;\right\rangle \;=\;U\;\left( \; \xi\;\right)  \;\left\vert \;\psi
_{0}\;\right\rangle \;\rightarrow\;\left\vert \;k,\;\xi\;\right\rangle\; = \;D\;\left(
\;\xi\;\right)\;  \left\vert \;k,0\;\right\rangle
\]%
\begin{align*}
\left\vert \;k,\,\zeta\;\right\rangle  &  = \; \;\exp\left( \,\xi\, K_{+}\;-\;\xi^{\ast}%
K_{-}\right)  \left\vert \;k,0 \;\right\rangle \\
&  = \;\;\left(\;  1\;-\;\left\vert \;\zeta\;\right\vert ^{2}\right)  ^{k}\;\exp\;\left( \, \zeta
K_{+}\,\right) \, \left\vert \;k,0 \;\;\right\rangle
\end{align*}
where $$\xi \;= \;-\;\left( \; \chi/2\;\right) \; e^{-i\,\varphi},\;\;\;\; \zeta \;= \;-\;\tanh\;\left(\;
\chi/2\;\right) \; e^{-i\,\varphi}\;=\;\frac{\xi}{\;\left\vert\; \xi\;\right\vert }%
\tanh \;\left\vert \;\xi\;\right\vert  $$ 

\medskip

The parameter $\zeta$  is restricted by
$\left\vert \;\zeta \;\right\vert\; <\;1 \;\left(disk \right)$, and the number $k$ is known as the Bargmann index which separates different irreducible representations. 

 $\zeta$ is taking to be a slowly function of time, as usual.

\medskip

Notice that the displacement operator $ D$ does not contain $K_0$. We will give a realization of the generators 
$K_{+},\, K_{-}$ and $K_{0}$ in the next Section.

\medskip

The state $\left\vert \;k,\;0\;\right\rangle $ meets the conditions: $K_{-}\;\left\vert
\;k,\;0\;\right\rangle\; = \;0$ and $K_{0}\;\left\vert\; k,\;0\;\right\rangle \;= \;k\left\vert
\;k,\;0\;\right\rangle .$

If one assumes the diagonal operator $K_{0}$ like the Hamiltonian we have that
$\left\vert \;k,\;\xi\;\right\rangle $ is eigenstate of the following Hamiltonian:%
\begin{align*}
\widetilde{H}  & \; = \;D\,\left( \; \xi\;\right) \; K_{0}\,\;D^{\dagger}\;\left(\;  \xi\;\right) \\
&  = \;\; K_{0}\;\cosh\;\chi\;+\;\frac{1}{2}\;\left( \; e^{-i\varphi}\;K_{+}\;+\;e^{i\varphi}%
\;K_{-}\;\right) \; \sinh\;\chi
\end{align*}
Taking into account $\zeta$ as a slowly varying function of the evolution
parameter of the system (e.g. time) and
\[
\left\langle \;k,\;\zeta\;\right.  \left\vert \;k,\;\zeta\;\right\rangle \;=\;\; 1
\]%
\[
\left\langle\; k,\;\zeta\;\right\vert K_{+}\left\vert\; k,\zeta\;\right\rangle
\,= \,\frac{2\,k\,\zeta^{\ast}}{\,\left(\;  1\;-\;\left\vert \;\zeta\;\right\vert ^{2}\;\right)
}\;=\;-\;k\;\sinh\,\chi\, e^{i\varphi}%
\]
then%
\[
\left\langle \;k,\;\zeta\;\right\vert \;\frac{d}{dt}\;\left\vert \;k,\;\zeta\;\right\rangle \;
= \;k\;\;\frac{\zeta^{\ast}\;\frac{d\zeta}{dt}\;-\;\zeta\;\frac{d\zeta^{\ast}}{dt}}{\left(\;
1\;-\;\left\vert \;\zeta\;\right\vert ^{2}\;\right)  }%
\]

\bigskip

From the definitions for the Berry phase, with $R\left(t\right)$ the set
of parameters (generally of geometrical origin) we have:%
\begin{align*}
\gamma\left( \, C\,\right)   &  \;=\;i\;\int_{0}^{T}dt\;\left\langle\; n,\;R\;\right\vert\;
\frac{d}{dt}\;\left\vert \;n,\,R\;\right\rangle \\
&  =\;i\;\int_{C}dR\;\left\langle \;n,\,R\;\right\vert \;\nabla_{R}\;\left\vert\;
n,\,R\;\right\rangle
\end{align*}

and the Berry phase expresses as%

\[
\gamma_{k}\left( \, C \,\right) \;\, =\;-\;i\;k\;\int_{C}\;\frac{\;\zeta^{\ast}\;d\zeta\;-\;\zeta\;
d\zeta^{\ast}}{\;\left( \; 1\;-\;\;\left\vert\; \zeta\;\right\vert ^{2}\;\right)  }%
\]

\bigskip

\bigskip

{\bf Mp(2), SU(1,1) and Sp(2):}

\bigskip
All the groups Mp(2), Sp(2,R), and SU(1,1) are three dimensional. It is
possible to parameterize them in several ways that make the homomorphic relations
particularly simple. We use two of such parameterizations, both of which are
described as:
$$ 
Mp\,(2)\;\rightarrow \;e^{-\,i\, \alpha_{1}T_{1}},\;e^{-\,i\, \alpha_{2}T_{2}},\; e^{-\,i \,\alpha
_{3}T_{3}}$$
\vspace{-0.5em}
$$Sp\,(2 {R}) \;\rightarrow \;\left(
\begin{array}
[c]{cc}%
e^{\frac{1}{2}\alpha_{1}} & 0\\
0 & e^{-\frac{1}{2}\alpha_{1}}%
\end{array}
\right)  ,\left(
\begin{array}
[c]{cc}%
\cosh\frac{1}{2}\alpha_{2} & \sinh\frac{1}{2}\alpha_{2}\\
\sinh\frac{1}{2}\alpha_{2} & \cosh\frac{1}{2}\alpha_{2}%
\end{array}
\right)  ,\left(
\begin{array}
[c]{cc}%
\cos\frac{1}{2}\alpha_{3} & - \sin\frac{1}{2}\alpha_{3} \\
\sin\frac{1}{2}\alpha_{3} & \cos\frac{1}{2}\alpha_{3}%
\end{array}
\right) 
$$

$$SU(1,1)\rightarrow\left(
\begin{array}
[c]{cc}%
\cosh\frac{1}{2}\alpha_{1} & \sinh\frac{1}{2}\alpha_{1}\\
\sinh\frac{1}{2}\alpha_{1} & \cosh\frac{1}{2}\alpha_{1}%
\end{array}
\right)  ,\left(
\begin{array}
[c]{cc}%
\cosh\frac{1}{2}\alpha_{2} & i\sinh\frac{1}{2}\alpha_{2}\\
-i\sinh\frac{1}{2}\alpha_{2} & \cosh\frac{1}{2}\alpha_{2}%
\end{array}
\right)  ,\left(
\begin{array}
[c]{cc}%
e^{\frac{i}{2}\alpha_{3}} & 0\\
0 & e^{-\frac{i}{2}\alpha_{3}}%
\end{array}
\right)$$

\bigskip

where the angle $\alpha_{3}$ has the range $(-4\pi,4\pi]$ for $Mp(2)$, and the range
$(-2\pi,2\pi]$ for $Sp\,(2,R)$ and $SU(1,1)$.

\bigskip

Let us to consider the brief description of the relevant symmetry group to
perform the realization of the Hamiltonian operator of the problem. This group
specifically is the Metaplectic $Mp\left(2\right)  $ as well as the groups that are
topologically covered by it. The generators of $Mp\left(2\right)  $ are the
following :
\begin{align}
T_{1}  &  \;\;= \;\frac{1}{4}\;\left(\,  q\,p\;+\;p\,q\;\right) \;\; = \;\;\frac{i}{4}\;\left(\,  a^{+2}%
-\;a^{2}\,\right)  ,\label{cr}\\
T_{2}  &  \;\;= \;\;\frac{1}{4}\;\left(\,p^{2}\;-\;q^{2}\,\right)  \;\;= \;-\;\frac{1}{4}\;\left(\,
a^{+2} + \;a^{2}\right)  ,\nonumber\\
T_{3}  & \; \;= \;-\;\frac{1}{4} \;\left(\,  p^{2}\;+\;q^{2}\,\right) \;\; = \;-\;\frac{1}{4}\;\left(
a^{+}a \; + \;a\,a^{+} \,\right) \nonumber
\end{align}
with the commutation relations,%
\[
\left[ \; T_{3},\;T_{1}\;\right]  \;=\;i\;T_{2},\;\;\;\;\left[ \; T_{3}, \;T_{2}\;\right] \; =\;-\;i\;T_{1}%
,\;\;\;\; \left[ \; T_{1}, \;T_{2}\;\right] \; = \;-\;i\;T_{3}%
\]
being $(q,\,p)$ ,(alternatively $(a,\,a^{+}))$ the variables of the standard harmonic
oscillator, as usual. 

\medskip

If we rewrite the commutation relations as: 
$$\left[\;
T_{3}\;,\;T_{1}\;\pm \;i\;T_{2}\;\right] \; = \;\pm \;\left(\;T_{1} \;\pm \; i\;T_{2} \;\right)  ,\;\;\;\left[\;
T_{1} \;+ \;i_;T_{2}\;, \;T_{1}\;-\;i\;T_{2}\;\right] \; = \;- \; 2\;T_{3}$$ \ 

we see that the states 
$\left\vert \;n \; \right\rangle $  are eigenstates of $T_{3}$:%
\[
T_{3}\;\left\vert \;n\;\right\rangle = \;-\;\frac{1}{2}\;\left(  n +\frac{1}{2}\right)
\left\vert \;n \;\right\rangle
\]
and it is easy to see that:\ $$T_{1}\;\;+\;\;i\,T_{2}\;=\;-\;\frac{i}{2}\;a^{2}, \;\;\;\;\; T_{1}
-\;i\;T_{2}\;\;=\;\;\frac{i}{2}\;a^{+2}.$$

\subsection{Quadratic Hamiltonians, the parametric oscillator and the group
structure}

The problem we are going to face: we know that the general parametric
oscillator has a quadratic structure of the known form%
\begin{equation}
\widehat{H}\;=\;\frac{1}{2}\;\left[\;Z\left(  t\right)\;  \widehat{p}^{2}\;+\;Y\left(
t\right) \; \left(\;  \widehat{p}\;\widehat{q}\;+\;\widehat{q}\;\widehat{p}\;\right) \;
+\;X\left(  t\right)\;  \widehat{q}^{2}\;\right]  \label{h1}%
\end{equation}
Therefore, using the group manifold, precisely the $Mp\left(  2\right)$ group, the
association is direct:%
\begin{equation}
\widehat{H}\;=\;-\;i\;\left[ \; Z\left(  t\right)  \widehat{K}_{+}\;+\;2\;Y\left(  t\right)\;
\widehat{K}_{0}\;+\;X\left(  t\right)  \widehat{K}_{-}\;\right]  \label{h2}%
\end{equation}
since, for Eq. (\ref{cr}): $$\widehat{K}_{0}\;=\;i\;T_{1},\;\;\;\;\widehat{K}_{+} = -\;i\;\left(
T_{2} + T_{3}\right)  ,\;\;\;\;\widehat{K}_{-} = \;i\;\left(  T_{2} \;-\; T_{3}\right)
\,,$$ consequently:
\begin{equation}
\left[  \widehat{K}_{+},\;\widehat{K}_{-}\right] \; = -\;2\;\widehat{K}_{0},\text{
\ \ \ \ }\left[  \widehat{K}_{0},\;\widehat{K}_{\pm}\right]  = \;\pm\;\widehat
{K}_{\pm} \label{cc1}%
\end{equation}

being then one of the possible representations, preserving the commutation
relations at the algebra level, the following:%

\begin{equation}
\widehat{K}_{0}=\left(
\begin{array}
[c]{cc}%
-1 & 0\\
0 & 1
\end{array}
\right)  ,\;\;\widehat{K}_{+}=\left(
\begin{array}
[c]{cc}%
0 & 0\\
1 & 0
\end{array}
\right),\;\;\widehat{K}_{-}=\left(
\begin{array}
[c]{cc}%
0 & 1\\
0 & 0
\end{array}
\right)  \rightarrow\widehat{H}=-\,i\left(
\begin{array}
[c]{cc}%
-Y\left(  t\right)  & X\left(  t\right) \\
Z\left(  t\right)  & Y\left(  t\right)
\end{array}
\right). \label{r2}%
\end{equation}

\medskip

Let us note that formally and for practical purposes we draw on a particular
parameterization of $SO(1,2)$, strictly according to the chain for
any element $A_{\alpha}$
\begin{equation}
A_{\alpha}:\;\;\in Mp\left(  2\right)  \;\supset \;Sp\left(  2\mathbb{R}\right)  \;\sim \;
SU(1,1)\;\supset\; SO(1,2)\;\approx \;L\left(  3\right)  \label{chain}%
\end{equation}
This is for the obvious reason that $Mp(2)$ does not have finite realizations,
being the double covering of $SU(1,1)$, $Sp(2R)$, and the quadruple covering
of $SO(1,2)$, properties to take into account when signs, phases and
connectivity become very important, as when determining the spectrum of
physical states of a particular system. 

\medskip

Consequently, we remark here that the
representation Eq.(\ref{r2}) is for practical purposes quite useful mainly at the
level of making a more precise comparison at the time to pick a point of
contact with other methods such as the one based on the Wei Norman theorem
\cite{WN}. Strictly speaking (we will return to this point at the end of this
paper) one must work independently of the representation as far as
non-compact groups are concerned (e.g. infinite dimensional unitary representations).

\subsection{Generalized parametric oscillator and the Berry phase}%

We take again as the starting point the following  Hamiltonian: 
\begin{equation}
H\left(  t\right) \; = \;Z \left(  t\right) \, \frac{p^{2}}{2m}\;+ \;\frac{\omega_{0}}%
{2}\;Y\left(  t\right) \, \left(\,  pq\;+\;qp\,\right)  \;+\;\epsilon \;X \left(  t\right)
\,\frac{m\left(  \omega_{0} \,q\,\right)  ^{2}}{2},\text{ \ \ \ }\left( \, \epsilon\,
= \,\pm1\,\right)  \label{h}%
\end{equation}

\medskip

Introducing as usual, annihilation-creation operators in  Eq.(\ref{h})
\begin{equation}
a \;= \;\frac{p\;-\;i\,m\;\omega_{0}\;q}{\sqrt{\;2\,m\;\omega_{0}\;\hbar}},\text{ \ \ \ \ \ }%
a^{+}\;=\;\frac{\;p\;+\;i\,m\;\omega_{0}\;q}{\sqrt{\;2\,m\;\omega_{0}\;\hbar}} \label{ac}%
\end{equation}
we obtain%
\begin{equation}
H\left(  t\right)  =\frac{1}{2}\left[  -\left(  \epsilon X\left(  t\right)
-Z\left(  t\right)  \right)  \left(  a^{+2}+a^{2}\right)  +iY\left(  t\right)
\left(  a^{+2}-a^{2}\right)  +\left(  \epsilon X\left(  t\right)  +Z\left(
t\right)  \right)  \left(\,aa^{+} + a^{+}a \, \right)  \right]  \label{ha}
 \end{equation}
where $\left(\epsilon \;= \;\pm \, 1 \;\right)$. We note that using a Bogoliubov type transformation like%
\begin{equation}
\left(
\begin{array}
[c]{c}%
b\\
b^{+}%
\end{array}
\right) \; = \;\left(
\begin{array}
[c]{cc}%
M_{+-} & \;\; M_{-+}\\
M_{--} & \;\; M_{++}%
\end{array}
\right)  \left(
\begin{array}
[c]{c}%
a\\
a^{+}%
\end{array}
\right)  \label{tr1}%
\end{equation}
where%
\begin{equation}
M_{\pm\pm} \;= \;\frac{1}{2\;\sqrt{\,\kappa\, Z}}\;\left[ \; \left( \, Z\;\pm \;\kappa\;\right) \; \pm\;
i\,Y\;\right]  \label{M1}%
\end{equation}
and
\begin{equation}
\kappa\;=\;\left(\,  \epsilon XZ-Y^{2}\,\right) ^{1/2} \label{ka}%
\end{equation}
the Hamiltonian takes the form%
\begin{equation}
H\;=\;\hbar\,\underset{\kappa\left(  t\right)  \omega_{0}}{\underbrace{\omega\left(
t\right)  }}\left( \; b^{+}b\;+\;\frac{1}{2}\;\right)  \label{hb}%
\end{equation}
due to the canonical commutation relations being preserved, e.g. $[\,b,\,b^{+}%
] \; = \; [\,a,\,a^{+}] = 1$,  and where we have used the relation $ (b^{+} \,b \; + \;  b\, b^{+})$ which follows from Eqs. (9) and (10):
\begin{equation}
b^{+}b \;\,+ \,\;b\,b^{+} = \frac{1}{4\kappa}\left[ \, 2\left(  Z\left(  t\right)  - \epsilon
X( t) \right)  \left(  a^{+2} + a^{2}\right)  + 2\left(  \epsilon
X\left(  t\right)  + Z\left(  t\right)  \right)  \left(  a\,a^{+} + a^{+}a\right)
+ 2\,iY(t) \left(  a^{2} - a^{+2}\right) \right]  \label{pr}%
\end{equation}

The Heisenberg equations of motion of the system are: %
\begin{align*}
\frac{d\,b}{d\,t}  &  \;=\; -\;\frac{i}{\hbar}\;\left[\;  b,\;H\;\right]  \;\;+\;\;\frac{\partial\,
b}{\partial\, t}\\
& \; =\;-\;i\;\omega\; b\;\;+\;\;\frac{i\;Z}{2\;\kappa}\;\left[ \; \frac{d}{dt}\;\left( \; \frac{Y}%
{Z}\;b\;\;-\;\;\frac{\;Y\;-\;i\;\kappa}{Z}\;b^{+}\;\right) \;\right]  ,
\end{align*}%

\begin{align*}
\frac{db^{+}}{dt}  & \; = \;-\;\frac{i}{\hbar}\;\left[\; b^{+},\; H\;\right]\; \; + \;\; \frac{\partial\,
b^{+}}{\partial\, t}\\
&  \;= \;i\;\omega\;b^{+}\;\;-\;\;\frac{i\;Z}{2\;\kappa}\;\left[ \; \frac{d}{dt}\;\left( \; \frac{Y}
{Z}\;b^{+}\;\;+\;\;\frac{Y\;+\;i\;\kappa}{Z}\;\;b\;\right) \;\right]  ,
\end{align*}

\medskip

Notice that for the purpose to find the Berry phase, the terms of interaction
between the creation and annihilation operators can be disregarded due that
are beyond the second order of adiabaticity. Consequently, the approximate
solution is given by

\begin{equation}
b\left(  t\right) \; \rightarrow \; b\left(
0\right) \;\exp {\left[\,-\,i\,\int\,\left[\,  \omega\,\left(  t\right)
\,-\,\frac{Z}{2\,\kappa}\,\frac{d}{dt}\,\left( \, \frac{Y}{Z}\,\right)  \,\right] \, dt\,\right]}  \label{ev}%
\end{equation}

\medskip

The second phase factor will be related with the Berry phase of the physical
system: e.g.: 
$$
B_{ph}\;\rightarrow \;\int  \,\frac{Z}{2\kappa}\;\frac{d}{dt}\left(\;
\frac{Y}{Z}\;\right)\;  dt.
$$

  For the sake of completeness, we have now computed the second order in adiabaticity for the Berry phase and 
 its full expression has the following interesting structure:
From the solution of the operators $b\left(  t\right)  $ up to second
order in adiabaticity, the Berry phase takes the form 

\[
B_{ph}\rightarrow\underset{1st\text{ }order}{\underbrace{\int\frac{Z}{2\;\kappa
}\;\frac{d}{dt}\left( \; \frac{Y}{Z}\;\right)  dt}}\underset{2nd\text{ }%
order}{\underbrace{\;-\int\frac{\hslash}{\kappa\;\omega_{0}}\;\left[\;  \frac{1}%
{2}\;\frac{d}{dt}\left( \; \frac{1}{Z}\frac{dZ}{dt}\;\right)  \;-\;\frac{1}{4}\left(\;
\frac{1}{Z}\;\frac{dZ}{dt}\;\right)^{2}\;\right]  }}%
\]
\\
The second term is the second 
order correction in $1/\hbar$ and as we can see the dependence is only on $Z(t)$, through \, $ d/dt\, [\,ln Z(t)\,]$ \, and its quadratic power, while the first order contains the truly parametric oscillator variable $Y(t)$. Clearly,  neglecting the 2nd order is fully justified. 

It is clear that adiabaticity applies in many space-times which are smooth, non singular, namely without sudden schock waves for instance. The  space times we are considering in the paper: de Sitter space-time and the regular black holes (with the quantum space-time  removed singularity) are widely justified here.

\subsection{Coherent state quantum evolution}%

As we have seen so far, from the point of view of the dynamics the relevant symmetries are dominated by the Metaplectic group and the groups covered by it, which define the symplectic and projective characteristics of the quantum phase space. Consequently, to illustrate the motion in the projective Hilbert space it is appropiate to start the corresponding coherent state given by

\begin{equation}
\left\vert \;\xi \;\right\rangle \;= \;e^{\,-\,\left\vert \,\xi \,\right\vert ^{2}} \;\underset
{n=0}{\overset{\infty}{%
{\displaystyle\sum}
}} \;\left\vert \;n \;\right\rangle \;= \;e^{\; (\;\xi \,b^{+2}\;\, - \,\;\xi^{\ast}b^{2}\;)}\;\left\vert
\;0 \;\right\rangle \label{comp}%
\end{equation}

\medskip

 The coherent state Eq.(16) comes from the operators \,$b^{+} \, b$  corresponding to the diagonal
Hamiltonian of equation (13) necessary for the analysis of the dynamics of the system.
 
\bigskip

\medskip

{\bf The Mp(2) Squeezed Vacuum and Physical States:}

\label{vacuumphys}

\bigskip

The displacement operator in the case of the vacuum squeezed state is an element of the
Mp(2) group written in the respective variables of the canonical annihilation and
creation operators.
\begin{equation}
S\left(  \,\xi\,\right)  \; = \;\exp \,\frac{1}{2}\left(  \;\xi^{\ast} \,b^{2}%
\;-\;\xi\, b^{+2}\;\right)  \;\;\in\;\;Mp\left(  2\;\right)  \label{s}%
\end{equation}
Seeing  Eqs. (\ref{tr1})-( \ref{ss}) the relationship is shown directly:%

\begin{equation}
\left(
\begin{array}
[c]{c}%
b\\
b^{+}%
\end{array}
\right)  \;\rightarrow\;S\left(  \,\xi\,\right)  \left(
\begin{array}
[c]{c}%
a\\
a^{+}%
\end{array}
\right)  S^{-1}\left(  \,\xi\,\right)  \;\,=\;\left(
\begin{array}
[c]{cc}%
\lambda & \mu\\
\mu^{\ast} & \lambda^{\ast}%
\end{array}
\right)  \left(
\begin{array}
[c]{c}%
a\\
a^{+}%
\end{array}
\right)  \label{ss}%
\end{equation}
\newline From Eqs (\ref{s}) (\ref{ss}), we see that the dynamics of these
"square root" fields of $\Phi_{\gamma}$, in the particular representation that
we are interested in, is determined by considering these fields as coherent
states in the sense that they are eigenstates of $a^{2}$ via the action of the
Mp$\left(  2\right)  $ group that is of the type:
\begin{align}
\left\vert \;\Psi_{1/4}\;\left(\,  0,\,\xi,\,q\,\right)  \;\right\rangle  &
\;=\;\; \overset{+\infty}{\underset{k=0}{\sum}}\; \;f_{2k}\,\left(  \,0,\xi\,\right)\;
\; \;\frac{\left(  a^{\dag}\right)  ^{2k}}{\sqrt{\,\left(  \,2k\,\right)  \,!}%
}\;\left\vert \;0\;\right\rangle \label{psi1/4}\\
\left\vert \;\Psi_{3/4}\;\left(  \,0,\,\xi,\,q \,\right)  \;\right\rangle  &
\;=\;\;\overset{+\infty}{\underset{k=0}{\sum}}\;\; f_{2k+1} \,\left( \,0, \,\xi\,\right)
\;\;\frac{\left(  a^{\dagger}\right)  ^{2k+1}}{\sqrt{\,\left(  \,2k+1\,\right)
\,!}}\;\left\vert \;0\;\right\rangle \label{psi3/4}%
\end{align}

\bigskip

For simplicity, we will take all normalization and fermionic dependence or
possible fermionic realization, into the functions $f\left(  \xi\right)  $.
Explicitly, at $t=0$, the states are:

\begin{equation}%
\begin{array}
[c]{c}%
\left\vert \,\Psi_{1/4}\;\left(  \;0,\,\xi,\,q\;\right)  \;\right\rangle
\;\;=\;\;f\left(  \,\xi\,\right)  \;\left\vert \;\alpha_{+}\;\right\rangle \\
\left\vert \,\Psi_{3/4}\,\left(  \,0,\,\xi,\,q\;\right)  \;\right\rangle
\;\;=\;\;f\left(  \,\xi\,\right)  \;\left\vert \;\alpha_{-}\;\right\rangle
\end{array}
\label{1/4-3/4}%
\end{equation}

\bigskip

where $\left\vert \alpha_{\pm}\right\rangle $\ are the coherent  basic states in the
subspaces \thinspace\ $\lambda\;=\;\frac{1}{4}$ \thinspace and \thinspace
\ $\lambda\;=\;\frac{3}{4}$ \thinspace\ of the full Hilbert space. In other
words, the action of an element of $Mp\left(  2\right)  \,$ keeps them
invariant (coherent), ensuring the irreducibility of such subspace, e.g :
\begin{equation}
\mathcal{H}\;\;\sim\;\;\left(
\begin{array}
[c]{cc}%
\mathcal{H}_{1/4} & \\
& \mathcal{H}_{3/4}%
\end{array}
\right)  \label{splitH}%
\end{equation}
Consequently, the two symmetric and antisymmetric combinations $(\pm)$ of the
two sets of states ${(1/4,\;3/4\,)}$ will span \textit{all} the Hilbert space:
$\mathcal{H}$:
\begin{equation}
\left\vert \,\Psi_{\pm}\;\right\rangle \;\;=\;\;\left\vert \,\Psi
_{1/4}\;\right\rangle \;\pm\;\left\vert \,\Psi_{3/4}\;\right\rangle
\;\;,\qquad\left\vert \;\pm\;\right\rangle \;\;=\;\;\left\vert
\;+\;\right\rangle \;\pm\;\left\vert \,-\,\right\rangle \label{completes}%
\end{equation}

The time development is obtained by applying Equation (\ref{ev}) to the
combined state (sweeping the entire $\mathcal{H)}$:
\[
\left.  \left\vert \xi\right\rangle \right.  _{t}=e^{-i\Gamma\left(
t,R\right)  }\left\vert \xi e^{+i\Gamma\left(  t,R\right)  }\right\rangle
\]
or, in the respective irreducible subspaces of $\mathcal{H}$ from the point of
view of $Mp(n)$:

\[%
\begin{array}
[c]{c}%
\left.  \left\vert \,\Psi_{1/4}\;\right\rangle \right.  _{t}\;=\;e^{-i\Gamma
\left(  t,R\right)  }\left\vert \,\Psi_{1/4}e^{+i\Gamma\left(  t,R\right)
}\right\rangle \\
\left.  \left\vert \,\Psi_{3/4}\,\;\right\rangle \right.  _{t}\;\;=e^{-i\Gamma
\left(  t,R\right)  }\left\vert \,\Psi_{3/4}e^{+i\Gamma\left(  t,R\right)
}\right\rangle
\end{array}
\]

\medskip

where the total phase (dynamical plus Berry phase) is%

\medskip

\[
\Gamma \left( \; t,\,R \,\left(  t\,\right)  \,\right)  \; = \; \underset{dyn.phase}%
{\underbrace{\int \,\omega\left(  t\right)\,  dt}}\underset{Berry\text{ }%
phase}{\underbrace{\;-\;%
{\displaystyle\int}
\;\frac{Z}{2\kappa} \;\frac{d}{dt}\left(  \frac{Y}{Z}\;\right) \; dt}}%
\]

\section{Relevant Applications in Cosmology }

New illustrative applications of this formalism appear in the context of cosmology and black holes. These situations correspond to the cases $Z(t) = X(t) = 1$.

\bigskip
 
In general, the relevant term for the emergence of the Berry phase is the linear 
part in the temporal differential equation of the parametric oscillator, or the $(qp + 
 pq)$ term in the Hamiltonian  Eq.(\ref{ha}) and thus the non vanishing coefficient $ Y (t)$. 
 
 \bigskip
 
 The diagonalization can be always performed through the Bogoliubov transformation Eq.(\ref{tr1}) - Eq.(\ref{pr}) to the quadratic Hamiltonian.  Therefore, for these cases we have:

\begin{equation}
H\left(  t\right) \; = \;\frac{1}{2}\left[\left(   
 1 -  \epsilon  \right)  \left(  a^{+2}+a^{2}\right) +iY\left(  t\right)
\left(  a^{+2}-a^{2}\right)  +\left( 1 + \epsilon \right)  \left(  aa^{+}+a^{+}a\right)  \right]
\end{equation}
\begin{equation}
M_{\pm\pm} \;=\; \frac{1}{2\sqrt{\kappa }} \left [ \;1\;\pm \;\kappa \; \pm \;
iY\;\right]   \;\;\;\;\;  \epsilon \;=\; \pm \;1 \label{M2}
\end{equation}

\medskip

$(\epsilon \;= \,- \,1 \,$ corresponding to the inverted, eg with imaginary frequency,  oscillator). 

And the Hamiltonian becomes
\begin{equation}
H \;=\; \hbar\; \omega \left( t \right) \left(  b^{+}b\; +\; \frac{1}{2}\right)  \label{hb2}%
\end{equation}
\begin{equation}
\kappa^2 = \left(\epsilon - Y^{2}\right), \;\;\;\; \;\;  \hbar \;\omega \left(
t\right) =  \kappa (t) \;\omega_0 \label{hb3}%
\end{equation}
\begin{equation}
b^{+}b + b\,b^{+}=\frac{1}{2\kappa}\;[\;\left( 1 -\epsilon
 \right)  \left(a^{2} + a^{+2}\right)  + 
i Y \left( t\right) \left( a^{2}-a^{+2}\right) +
 \left( 1 + \epsilon
\right)  + \frac{1}{2} \left(aa^{+} + a^{+}a\right)
 \;]  \label{pr2}%
\end{equation}

\medskip

The Berry phase becoming
\begin{equation}
B_{phase} = \int\frac{dt}{2\, \sqrt{ \;\epsilon - Y^{2}\;}}\;\;\frac{d}{dt} \;Y \;  
\end{equation}

\medskip

\subsection{Berry Phase of de Sitter Inflation}

 \bigskip
 
 Interestingly enough, de Sitter space-time can be described as an {\it inverted} harmonic oscillator, eg with imaginary frequency,  both classically and at the quantum level, the  Hamiltonian taking the same form of Eq. (\ref{hb2}) with 

$$  |\kappa|^2 = (1 + Y^2). $$

The description of de Sitter inflation is preciselly that of a parametric oscillator with

$$  Y (t) =  H(t) = \sqrt {\;\Lambda (t) \,/\,3\;} ,$$ 

the oscillator length being $ l_{osc} =  1/H$. The Berry phase 
of de Sitter inflation is in fact imaginary, and with the sign describing the increasing exponential factor acceleration

\begin{equation}
B_{\,phase}\, = - \,i \;\int_{t_{in}}^{t_{end}} \;\frac{dt}{2\; \sqrt{ \;1 \;+ \;H^2 \;}}\;\;d_t\, H 
\end{equation}

\medskip

This integral is explicitly solved as a function of $H$ with the limits: $$H(t_{end}) \equiv H_{e} , \;\;\; H(t_{in})\equiv H_{i}:$$

\begin{equation} \label{binf1}
B_{phase}\; = \; \frac{\,i\,}{\,2\,}\;\ln \;\left (\;\frac{\;H_{i}\;+\;\sqrt{\;1\;+\;H_{i}^{2}\;}}{\;H_{e}\;+\;\sqrt{\;1\;+\;H_{e}^{2}\;}\;}\;\right)%
\end{equation}

\bigskip

The finite interval $[\, t_{in} \,, \,t_{end} \,]$ indicating the initial and the final time of inflation. 

\medskip

 In the early universe inflationary phase, the Hubble values are both very high: $ ( H_i , H_e ) >> 1$, typically of the Grand Unification scale, and therefore the expression for the Berry phase Eq.(\ref{binf1}) 
yields in the early universe:

\begin{equation} \label{binf2}
B_{phase}\; = \; \frac{\;i\;}{\;2\;} \;\ln\, \left( \frac{\;H_i\;}{\;H_{e}\;}\right) \; + \; \frac{\;i\;}{\;2\;}\;\left(\;\frac{\,1\,}{\,H_i^2\,}\; - \; \frac{\,1\,}{\,H_e^2\,} \;\right) \;+ \;O \;\left(\frac{\;1\;}{\; H_i^2\;H_{e}^2\;}\right)
\end{equation} 

\bigskip

Interestingly, the Berry phase for de Sitter inflation expresses as the logarithm of the quotient between the initial and final values of the Hubble constant during inflation. The imaginary value of the Berry phase here appropriately corresponds to the exponential expanding nature of the de Sitter background.

\subsection {Berry Phase of the Inflationary Perturbations:}  

\medskip

The linear quantum perturbations of Inflation, both the scalar $(S)$ curvature fluctuations and the tensor $(T)$ gravitational $k$ - modes both satisfy a second order Schröedinger type equation in time:
\begin{equation}
Q^{\,'\,'}(\,\eta\,)_{\;(S,\,T)}\; + \;[\; k^2 \;-\; W^2\, (\,\eta\,)_{\;(S,\,T)}\;] \;Q (\,\eta\,)_{\;(S,\,T)} \;\; = \;\;0
\end{equation}\label{Q}
where $\eta$ is the conformal time related to the cosmic time $t$ by $ dt =  {a} (\eta) \,d \eta $,  
primes $('')$ denote the second  derivative with respect to $\eta$, and the potential $W^2\, (\,\eta\,)_{\;(\,S,\, T)}$ felt by the fluctuations is:
\begin{equation} \label{WS}
W^2\, (\,\eta \,)_{\;(\,S)}\; \;= \;\;  {a}^{\,'\,'} (\,\eta\,)\, /\,{a} (\,\eta\,) 
\end{equation}
\begin{equation} \label{WT}
W^2\, (\,\eta\,)_{\;(\,T)}\;\; =\;\;  z^{\,'\,'} (\,\eta\,)\, / \,z (\,\eta\,) 
\end{equation}
The variable $z$ appropriately combines the inflaton field $\phi (t) $ and the accelerated expansion background  
described by the scale factor $ a (t)$ driven by $\phi (t)$ ,  the Hubble parameter being $H \;= \; {a}^{.}(t)\;/{a}(t)$\,:
\begin{equation} \label{z1}
z \; \;= \;\; {a} (t) \;\; \frac{\;\phi^{\,.}(t)}{\;H}
\end{equation}
\begin{equation} \label{z2}
\frac{d^2 z}{d \eta^2} \;\; = \; \; {a^2}\; 
(\, z^{..}\; + \; H \;z^{.} \,)
\, \;\; 
\end{equation}

Therefore, the scalar and tensor gravitational inflationary fluctuations both satisfy similar parametric harmonic oscillator equations with the Hamiltonian  $\mathcal{H}$:
\begin{equation} \label{HH}
\mathcal{H} \; = \;\frac{1}{2}\;[\;   P^2 \; + \; \Omega^2\, (\,\eta\,)_{\;(S,\,T)}\,Q^2 \;]
\end{equation}
$$ \Omega^2\, (\,\eta\,)_{ \;(S,\,T)} \; \equiv \; k^2 \;-\; W^2\, (\,\eta\,)_{\;(S,\,T)}  $$ 
 
The two typical situations do appear:
$$
\text{ $\Omega\, (\,\eta\,)_{\;(S,\,T)}$\; \;Real \; :   \;\;\,$ k^2 \; > \;\; W^2\, (\,\eta\,)_{\;(\,S,\, T)}$, \; (sub horizon k-modes)}  
$$
$$
\text{ $\Omega\, (\,\eta\,)_{\;(S,\,T)}$ \, Imaginary : \; $ k^2 \; < \;\; W^2\, (\,\eta\,)_{\;(\,S,\, T)}$, \; (super horizon k-modes)}  
$$

\medskip

 Generically, the field oscillations for which the wavelengths $\lambda  = 1/k$ are inside the Hubble radius $1/H$  are named sub horizon modes, therefore $ k > H $ for them. The super horizon modes are those for which the wavelengths are larger than the Hubble horizon: $\lambda > 1/H $, namely  $ k < H $. Therefore, the above two regimes on the $k$ - modes of the inflationary fluctuations (in conformal time) precisely  correspond to these two typical situations. 

\bigskip

$\mathcal{H} (P,Q)$ correspond to the $(b, b^+)$ representation, 
and a canonical (Bogoliubov) transformation as Eq.(\ref{M2}):    $(P,Q) \rightarrow (q,p)$, [ equivalently  $(b, b^+) \rightarrow (a, a^+)$\,], yields $\mathcal{H}$ into the form: 
\begin{equation} \label{Hpq}
\mathcal{H} \; = \;\frac{1}{2}\;[\; p^2 \; + \; Y\, (\,\eta\,)_{\;(S,\,T)}\;(\,p\,q\;+\;q\,p\,) \;+\; k^2 \;q^2\;]
\end{equation}

$$ Y\, (\,\eta\,)_{\;(S)} \; = \; z' \,/\,z,  \;\;\;\;\; \;\; Y (\,\eta\,)_{\;(T)} \; = \; {a}' /\,{a} $$ 
 
with 
\begin{equation}
\Omega^2\, (\,\eta\,)_{ \;(S,\,T)} \;= \; (\,k^2 \;-\; Y\, _{ \;(S,\,T)}^2\,)\;-\; Y\,_{ \;(S,\,T)}'\; = \; \omega ^2 \;-\; Y\, _{ \;(S,\,T)}'
\end{equation}

\medskip

$ Y_{\;(S,\,T)}' \,= \, a" / a  \,- \, Y^2 $,  and recall here that  $\eta$ is the conformal time, $(') \equiv \;d/ d \eta$

\medskip

The Berry phase of the quantum inflationary fluctuations ${\;(S,\,T)} $ is then:

\begin{equation}\label{Binfk}
B_{\,phase\,,\,k } {\;(S,\,T)}  \;= \;\int_{\,0}^{\,\eta_{\,end}}\frac{d\eta}{2\; \sqrt{ \;k^2 \,- \,Y^{2}_{\;(S,\,T)}\;}}\;\;Y_{\;(S,\,T)}'\;  
\end{equation}

\bigskip

The $\eta$ interval in the integral depends on whether it refers to sub horizon or super horizons modes, eg on whether \; $k^2 \,> \,Y^2 (\,\eta_{\,end})_{\;(S,\,T)}$ \; or \;$k^2 \,< \,Y^2 (\,\eta_{\,end})_{\;(S,\,T)}$ respectively, \;$\eta_{\,end} $ \; being then fixed by:
$$  k^2 \; = \; Y^2 (\,\eta_{\,end}).   $$

Note that in the case of the cosmological perturbations the integral is explicitly
solved as a function of $Y_{\left( S,T\right) }(\eta )$ with limits $%
Y_{\left( S,T\right) }(0)\equiv Y_{0}$ and $Y_{\left( S,T\right) }(\eta
_{end})\equiv Y_{e}$:

\medskip

(i) sub horizon modes:%
\[
B_{phase},\text{ }_{k}\left( S,T\right) \;=\;\frac{1}{2}\;\arctan \left[\; \frac{%
Y_{e}-Y_{0}}{\sqrt{\;\left( k^{2}-Y_{e}^{2}\;\right) \left(
k^{2}-Y_{0}^{2}\;\right) }+Y_{e}Y_{0}}\;\right] 
\]

\bigskip

(ii) super horizon modes:%
\[
B_{phase},\text{ }_{k}\left( S,T\right) \;=\;\frac{\;i\;}{\;2\;}\;\ln \;\,\left(\;\frac{Y_{e}\;+\;\sqrt{\;
Y_{e}^{2}\;-\;k^{2}}}{Y_{0}\;+\;\sqrt{\;Y_{0}^{2}\;-\;k^{2}}}\;\right)
\]

\bigskip

We can also express the Berry phase  in the inflationary slow roll regime which is well appropriated here  because of the Berry phase adiabaticity. $Y (\,\eta)_{\;(S,\,T)}$ can be written in terms of the slow roll parameters $(\epsilon_v ,\;\eta_v)$, and therefore related to the cosmological observables: the scalar and tensor spectral indices $ n_S $ and $n_T$, and the scalar to tensor ratio $r$:

\begin{equation}
W\, (\,\eta\,)_{\;(S,\,T)} \;\; = \;\; 
\frac{(\,\nu^{2}_{\;(S,\,T)} \; - \; 1/4 \;)}{\eta^2}
\end{equation}
$$\nu_{\;(S)} \; = \;\; 3/2 \; + \; 3\,\epsilon_v \;- \;\eta_v \;\; = \;\; 3/2 \;-\;(\,n_s \;- \;1\,)\,/\,2 $$
$$ \nu_{\;(T)} \; \;= \;\; 3/2 \;+\; \epsilon_v \; = \; 3/2 \;-\; n_T \,/\, 2 $$
\begin{equation}
 r \; \; =\;\;16\; \epsilon_v\; = \;- \,n_T /\,8
\end{equation}

These expressions show the explicit  operational  applications of our framework. The analysis of these findings, their properties and  
the observable features in terms of real cosmological data  exceeds the purpose of this paper and deserve future work.

\section{Entanglement}

\subsection{Density matrix viewpoint}

 We will develop now the entanglement of quantum states in the theoretical
context of non-compact groups, in particular for the Metaplectic group.  Besides its several local and global interesting properties, the Metapletic group is strongly (mathematically and physically) supported by the principle  of minimum group representation described in detail in Ref. [2].

 \bigskip

Let us consider the following state%
\[
\left\vert \,\Phi\;\right\rangle \;=\;\left\vert \,\Phi_{A}\;\right\rangle
\left\vert \,\Phi_{B}\;\right\rangle
\]
The sub-indices A and B indicate the corresponding sub- systems. Usually one could even
define (in the context of our previous works) the density matrix for an
observable and an unobservable sector.%
\[
\left\vert \,\Phi\;\right\rangle \;= \;\left\vert \,\Phi o\;\right\rangle
\left\vert \,\Phi d\;\right\rangle
\]
in such a way that the density matrix of the observable system is%
\[
\rho_{o}\;=\;Tr_{d}\;\left\vert \,\Phi\;\right\rangle \,\left\langle \Phi
\;\right\vert
\]
and the corresponding entropy of the entanglement in this case%
\[
S_{E}\;=\;-\;Tr_{o}\;\rho_{o}\;\log\rho_{o}%
\]
Using the Schmidt decomposition, some pure state can be written as%
\begin{equation}
\left\vert \,\Phi\;\right\rangle\; = \;\overset{n}{\underset{i=1}{\;\sum}\;c_{i}%
}\;\left\vert \,v_{iA}\;\right\rangle\; \otimes\;\left\vert \,u_{iB}\;\right\rangle
\label{sch}%
\end{equation}
where $\,\left\vert \,v_{iA}\;\right\rangle ,\left\vert \,u_{iB}\;\right\rangle
\;$ are orthonormal states in the subsystems A and B respectively. If we
see the structure of the Hilbert space Eq. (\ref{splitH}) for a state $\in SU(1,1)$
the correspondence between Eq.(\ref{sch}) and the results of our previous papers
is immediate as we will demonstrate below.

Not all states are separable states (and thus product states). Fix a basis for
$H_{A}$ and a basis for $H_{B}$ subsystems.The most general state in
$H_{A}\otimes H_{B}$ is now of the form

.%
\[
\left\vert \,\Phi\;\right\rangle \;=\;\underset{i,j\,=\,1}{\sum}c_{ij}\;\left\vert
\,v_{iA}\;\right\rangle \otimes\left\vert \,u_{jB}\;\right\rangle
\]

This state is separable if there exist vectors such that yielding $c_{ij}%
= c_{iA}\,c_{jB}/\left\vert \,\Phi\;\right\rangle =\underset{i=1}{\sum}%
c_{iA}\,\left\vert \,v_{iA}\;\right\rangle \otimes\underset{j=1}{\sum}%
\,c_{jB}\,\left\vert \,u_{jB}\;\right\rangle$. And it is {\it inseparable} if for any
vectors at least for one pair of coordinates $c_{ij} \neq c_{iA}\,c_{jB}$ . If a
state is {\it inseparable}, it is called an {\bf 'entangled state'}. The typical case is
one of the Bell states e.g.%

\[
\frac{\left\vert \,1_{A}\;\right\rangle \otimes\left\vert \,0_{B}%
\;\right\rangle -\left\vert \,0_{A}\;\right\rangle \otimes\left\vert
\,1_{B}\;\right\rangle }{\sqrt{2}}%
\]

\medskip

of the four Bell states, which are (maximally) entangled pure states with the basis \; $\left\{  \left\vert
\,0_{A}\;\right\rangle ,\left\vert \,1_{A}\;\right\rangle \right\}  \in H_{A}$ \;
and \; $\left\{  \left\vert \,0_{B}\;\right\rangle ,\left\vert \,1_{B}%
\;\right\rangle \right\}  \in H_{B}$. (These are pure states
of the $H_{A}\otimes H_{B}$ space, but which cannot be separated into pure
states of each $H_{A}$ and $H_{B}$ space). 
If the composite system is in such state,
it is impossible to attribute to either system A or system B a definite pure
state. 

Notice that while the von Neumann entropy of the whole entangled state is zero (as
it is for any pure state), the entropy of the subsystems is greater than zero. In this sense, the systems are "entangled".

\medskip

Recently, in previous recent works \cite{cirilo-sanchez}, \cite{universe}, we have seen that physical states are mappings of the group generators in a vector representation, and that these are expressed in the so called Sudarshan's
diagonal-representation and in a \textit{new non diagonal} one that leads, as an important consequence, the
\textit{physical states} with spin content $\lambda
\;=\;(\,1/2,\;1,\;3/2,\;2\,)$.

Precisely, the {\it generalized coherent states} here
generate a map that relates the metric $g_{ab}$, solution of the wave equation, 
to the specific subspace of the full Hilbert space where these coherent
states live. Moreover, for operators $\in Mp\left(  2\right)
$ there exists an asymmetric - kernel leading for our case the following $\lambda = 1$ state :%

\begin{equation}
\left.  g_{ab}\left(  \,t,\,1,\,\alpha\right)  \,\right\vert _{HW}%
\;=\;\left\langle \Psi_{3/4}\left(  t\right)  \right\vert L_{ab}\left\vert
\Psi_{1/4}\left(  t\right)  \right\rangle =\;\mathcal{F}\;\left(
\begin{array}
[c]{c}%
\alpha\\
\alpha^{\ast}%
\end{array}
\right)  _{\left(  1\right)  \,ab} \label{11}%
\end{equation}

where
\[
\mathcal{F}\;=\;e\,^{[\,-\,\left(  \,\frac{m}{\sqrt{\,2}\;\left\vert
\,\mathbf{a}\,\right\vert }\,\right)^{2}\,\left[  \;\left(  \,\alpha
\,+\,\alpha^{\ast}\,\right)  \,-\,B\;\right]  ^{2}\,+\,D\;]}\,e\,^{[\;\xi
\,\varrho\;\left(  \,\alpha\,+\,\alpha^{\ast}\,\right)  \;]}\;\left\vert
\,f\left(  \,\xi\,\right)  \,\right\vert ^{2}%
\]
where $B$ and $D$ are given by:
\begin{equation}
B\;=\;\left(  \,\frac{\left\vert \,\mathbf{a}\right\vert }{m}\,\right)
^{2}\,c_{1},\quad\;  \;\quad D\;=\;\left(  \frac{\left\vert \,\mathbf{a}%
\,\right\vert \,c_{1}}{\sqrt{2}\,m}\,\right)  ^{2}\,+\,c_{2} \label{l}%
\end{equation}

\medskip

$c_{1}$ and $c_{2}$ being constants characterizing the solution or its initial
conditions; $\;\xi$ is the fermionic super coordinate of the
corresponding group manifold and $\varrho$ is the fermionic part of the superfield solution.

This is so because the non-diagonal projector involved in the reconstruction
formula of $L_{ab}$ is formed with the $\Psi_{1/4}$ and $\Psi_{3/4}$ states
which span completely the \textit{full} Hilbert space. And this is precisely a
particular case of the Schmidt decomposition of an SU(1,1) quantum state, namely%

\begin{equation}
\Phi_{ab}\equiv\underset{Mp\left(  2\right)  representation}{\underbrace
{\left\langle \Psi_{3/4}\left(  t\right)  \right\vert L_{ab}\left\vert
\Psi_{1/4}\left(  t\right)  \right\rangle }}\sim\left\langle \Psi_{3/4}\left(
t\right)  \right\vert \left(
\begin{array}
[c]{c}%
a\\
a^{+}%
\end{array}
\right)  _{\,ab}\left\vert \Psi_{1/4}\left(  t\right)  \right\rangle
\rightarrow\underset{Schmidt-repres.}{\underbrace{\,\Phi_{ab}\;=\overset
{}{\underset{n,m=0}{\sum}c_{ab}}v_{m1/4}\;\otimes u_{n3/4}^{\ast}\;}}
\label{12}%
\end{equation}

The above correspondence is fully consistent due to the fact that $\left\vert \Psi
_{1/4}\left(  t\right)  \right\rangle ,\left\vert \Psi_{3/4}\left(  t\right)
\right\rangle $ are pure states (coherent) in each subspace $\in SU(1,1)\ $ and they are
mutually orthogonal. 

With more detaill we have in the $Mp\left(  n\right)  $
case from Eq. $\left(  \ref{12}\right)  $:

\begin{align}
\Phi_{ab}\;  &  =\underset{n=0,1,2..}{\sum}\underset{u_{3/4}^{\ast}%
}{\underbrace{\left\langle \Psi_{3/4}\left(  t\right)  \right\vert \left.
2n+1\right\rangle }\text{ \ \ }}\underset{c_{ab}}{\underbrace{\left\langle
2n+1\right\vert L_{ab}\left\vert 2n\right\rangle }}\text{\ \ }\underset
{v_{1/4}}{\underbrace{\left\langle 2n\right.  \left\vert \Psi_{1/4}\left(
t\right)  \right\rangle }}\label{12a}\\
&  =\underset{n = 0,1,2..}{\sum}\left(  1-\left\vert \omega\right\vert
^{2}\right)  \frac{\left(  \omega^{\ast}/2\right)  ^{2n+1}}{\sqrt{\left(
2n+1\right)  !}}\left(
\begin{array}
[c]{c}%
0\\
\sqrt{2n+1}%
\end{array}
\right)_{\,ab}\frac{\left(  \omega/2\right)^{2n}}{\sqrt{\left(2n\right)
!}}=\nonumber\\
&  =\underset{n = 0,1,2..}{\sum}\Xi\left(  n\right)  \left(
\begin{array}
[c]{c}%
0\\
\sqrt{2n+1}%
\end{array}
\right)  _{\,ab}\omega^{\ast}\nonumber
\end{align}
Here we have used the discrete basis (series)\ in the disk of Mp(2) and its coverings.

Meanwhile the case%

\begin{align}
\Psi_{ab}  &  \equiv\left\langle \Psi_{1/4}\left(  t\right)  \right\vert
L_{ab}\left\vert \Psi_{3/4}\left(  t\right)  \right\rangle \neq\left\langle
\Psi_{3/4}\left(  t\right)  \right\vert L_{ab}\left\vert \Psi_{1/4}\left(
t\right)  \right\rangle \label{12b}\\
&  =\underset{n=0,1,2..}{\sum}\underset{u_{3/4}^{\ast}}{\underbrace
{\left\langle \Psi_{1/4}\left(  t\right)  \right\vert \left.  2n\right\rangle
}\text{ \ \ }}\underset{c_{ab}}{\underbrace{\left\langle 2n\right\vert
L_{ab}\left\vert 2n+1\right\rangle }}\text{\ \ }\underset{v_{1/4}}%
{\underbrace{\left\langle 2n+1\right.  \left\vert \Psi_{3/4}\left(  t\right)
\right\rangle }}\nonumber\\
&  =\underset{n=0,1,2..}{\sum}\Xi\left(  n\right)  \left(
\begin{array}
[c]{c}%
\sqrt{2n}\\
0
\end{array}
\right)_{\,ab}\omega\nonumber
\end{align}

with the definition%
\[
\Xi\left(  n\right) \; = \;\left(  \frac{\left\vert \omega\right\vert }{2}\right)
^{2n}\frac{\left(  1-\left\vert \omega\right\vert ^{2}\right)  }%
{2\sqrt{\left(  2n\right)  !\left(  2n+1\right)  !}}%
\]
\\
We can see that the expressions Eq.(\ref{12a})\ and Eq.(\ref{12b})\ describe two
definite chirality states showing the non compact analog to the SU(2)\ case with the
respective spinors (discrete series inside the disc with characteristic
complex variable $\omega$).\ 

It is also important to remark the following points:
 
 \bigskip

(i) We have for Mp(2) a $\infty-$dimensional but unitary representation that
maps it to the Lorentz group in 3-dimensional $L_{3}$.

\bigskip

(ii)  We can see that :%
\[
\left\langle \Psi_{3/4}\left(  t\right)  \right\vert L_{ab}\left\vert
\Psi_{1/4}\left(  t\right)  \right\rangle \neq\left\langle \Psi_{1/4}\left(
t\right)  \right\vert L_{ab}\left\vert \Psi_{3/4}\left(  t\right)
\right\rangle
\]

\medskip

This fact have also interesting implications, as for the realization of the CPT invariance and for the arrow of time. This is so because of the full covering of the Hilbert space by the $\left\vert 1/4\right\rangle $
 and  $\left\vert 3/4 \right\rangle $ sets of states ({\it even} and {\it odd} sets) of the Metaplectic (Mp) group. In particular, the correspondence between the standard Schrodinger cat states and the basic states  $\left\vert 1/4\right\rangle $
 and  $\left\vert 3/4 \right\rangle $ of Mp(2) is discussed in Section VII.

\bigskip

(iii) From the point (ii) it can be seen that there would be a natural transition
from the compact spinorial description of SU(2) for example, to the non-compact
SU(1,1) description by taking as the basis of this transition the universal covering Mp(2).

\bigskip

Let us also
notice that the bilinear state associated with spin 2 in this basis, similarly as
in the expressions Eqs.(\ref{12a}-\ref{12b}), was associated with the spacetime
metric, providing the space-time discretization condition for small $n$ and  the 
continuous spacetime for $n$ $\rightarrow\infty$.

Consequently, and due to Eq. (\ref{11}) the entanglement entropy is given by %

\[
S_{E} = - Tr \;\left\vert \; \mathcal{F}\,\alpha\;\right\vert ^{2}\;\log\;\left\vert\;
\mathcal{F}\,\alpha\;\right\vert^{2}%
\]
where $\mathcal{F}$ is 
due to expression Eq.(\ref{11}).

\section{Quantum Evolution, Adiabatic Invariants and Topological Structure}

Now, we will provide the interpretation of the relation between the coherent
states in the metaplectic representation and the quantum evolution of the
spacetime. For this purpose we will need to start similarly to Eq. (\ref{ss}) with the
association
\[
\left(
\begin{array}
[c]{c}%
b\\
b^{+}%
\end{array}
\right)  \rightarrow\left(
\begin{array}
[c]{c}%
a_{(+)}\\
a_{(+)}^{+}%
\end{array}
\right) \; \equiv \;L_{\left( +\right)  }%
\]
and%
\[
\left(
\begin{array}
[c]{c}%
a\\
a^{+}%
\end{array}
\right)  \rightarrow\left(
\begin{array}
[c]{c}%
a_{(-)}\\
a_{(-)}^{+}%
\end{array}
\right) \; \equiv \;L_{\left( -\right)  }%
\]

where $a_{(\pm)}$ correspond to operators $a\left(  t\right)  $ for 
$t\rightarrow\pm\infty$ (asymptotical values). Consecuently
\[
a_{(-)}^{+}a_{(-)}\left\vert n\left(  -\right)  \right\rangle = n\left\vert
n\left(  -\right)  \right\rangle , \;\; a_{(+)}^{+}a_{(+)}\left\vert n\left(+\right)  \right\rangle = n\left\vert
n\left(+\right)  \right\rangle, \;etc.
\]
From Eq.(\ref{ss}) we have
\begin{align}
L_{\left(  +\right)  }  &  =UL_{\left(  -\right)  }U^{-1}=\mathbb{A}L_{\left(
-\right)  },\text{ \ }\label{tr}\\
\text{\ \ \ \ \ }\mathbb{A}  &  \mathbb{\in}SU\left(  1,1\right),
\text{\ \ }\nonumber
\end{align}
but now the element of the group is%
\[
U\in Mp\left(  2\right)  /U \;= \;e^{-i\gamma \, T_{3\left(  -\right)  }}\;e^{-i\beta \,
T_{1\left(  -\right)  }}\;e^{-i\alpha \, T_{3\left(  -\right)  }}%
\]
We can observe now that
\[
M_{mn}=\left\langle m\left(  +\right)  \right\vert \left\vert n\left(
-\right)  \right\rangle
\]
where as before $a_{(+)}^{+}a_{(+)}\left\vert m\left(  +\right)  \right\rangle
= m\left\vert m\left(  +\right)  \right\rangle$. We can see also that:
$a_{(+)}^{+}a_{(+)} = U\, a_{(-)}^{+}\, a_{(-)}\, U^{-1}.$ From Eq. (\ref{tr}) :%
\[
\left\vert \,m\left(  +\right)\,  \right\rangle \;= \;U\;\left\vert \,m\,\left(  -\right)\,
\right\rangle
\]
then,%
\[
\left\langle m\left(  +\right)  \;\right\vert \left\vert n\left(  -\right)
\right\rangle \;= \;\left\langle m\left(  -\right)  \right\vert \;U^{-1}\;\left\vert
n\left(  -\right)  \right\rangle
\]
with our matrix element
\[
M_{mn} = \left\langle \;m\right\vert \;e^{i\alpha T_{3-}}\;^{i\beta T_{1}}\,e^{i\gamma
T_{3}}\;\left\vert n\right\rangle
\]
where the index $\left(  -\right)  $ was dropped for simplicity. Because
$T_{3}\left\vert n\right\rangle =-\frac{1}{2}\left(  n+\frac{1}{2}\right)
\left\vert n\right\rangle$ : %
\[
M_{mn}\; =\; e^{-i\alpha\frac{\left(  2m+1\right) }{4}}\;e^{-i\gamma\frac{\left(
2n+1\right)  }{4}}\left\langle m\right\vert\; e^{i\beta T_{1}}\;\left\vert
n\right\rangle
\]
We can further consider the following quantity that  determines the transition probability of the oscillator from the state
$\left\vert\, n\left(-\right)  \,\right\rangle $ to the state $\left\vert \,
m\left(+\right)  \,\right\rangle$ :%
\begin{equation*}
W_{mn}\; = \;\left\vert \;M_{mn}\;\right\vert ^{2}
 \; = \; \left\vert \;\left\langle \;m \;\right\vert e^{\,-\frac{\beta}{4}\,\left(\,
a^{+2}-\;a^{2}\,\right)  }\left\vert \;n\;\right\rangle \;\right\vert ^{2}%
\end{equation*}
In particular,%
\begin{align*}
W_{nn}  &  = \;\left\vert \;\left\langle \,n \,\right\vert e^{-\frac{\beta}{4}\,\left(\,
a^{+2} - \;a^{2}\,\right)  }\left\vert \,n\,\right\rangle \,\right\vert^{2}\\
&  = \;\left\vert \;\left\langle \, n \,\right\vert 1 \;- \;\frac{\beta}{4} \left(  a^{+2}%
-\;a^{2}\right)\;  + \;\frac{\beta}{32}^{2}\left(  a^{+2} -\;a^{2}\right)
^{2} -\;\cdot\cdot\cdot\;\left\vert n\right\rangle \;\right\vert ^{2}\\
&  = \;1-\;\frac{\beta}{8}^{2}\left( \; n^{2} \;+ \;n+1\;\right)^{2}\;+\;\cdot\cdot\cdot
\end{align*}

By taking $\beta < 1$, $\beta^{2}$ {\bf in  the above expression} is the magnitude
determining the accuracy of preservation of the adiabatic invariant of the
classical oscillator, generally with time-dependent frequency, (parametric oscillator), as
Eq.(\ref{h2}) related with the dynamical and geometrical phases. 

Let us notice that in the transition 
probability $W_{mn}$ the relevant generator 
of the transition between the eigenstates $\left\vert \,n \,\right\rangle $ of 
$T_3$ of the respective system is 
$T_1 $ characteristic of the
diabatic-adiabatic evolution of the physical system considered.

\section{Entanglement with Semi-Coherent States}

Let us briefly analyze in an algebraic description, the origin of the quantum
relativistic effects as the prolongated highly oscillations effect or so
called "Zitterbewegung". 

\medskip

There are two types of states in the
\ "algebro-pseudo-differential" correspondence: the basic
(non-observable) states and the observable physical states. In that case, the
basic states are coherent states corresponding to the double covering of the 
$SL(2C)$ group, eg the Metaplectic group \cite{cirilo-sanchez}, \cite{universe}: This is the responsible for
projecting the symmetries of the 6 dimensional $Mp(4)$ group space to the 4
dimensional space-time by means of a bilinear combination of the $Mp(4)$
generators. The supermultiplet metric solution for the geometric Lagrangian is
\[
g_{ab}(t,\lambda)\;=\;\left\langle \psi_{\lambda}\left(  t\right)
\;\right\vert \;L_{ab}\;\left\vert \;\psi_{\lambda}\left(  t\right)
\;\right\rangle
\]
As we can see above, the physical state (which appears as a mapping of the
non-compact generator of interest and its fundamental coverings) takes
precisely the form of a Husimi quasi-probability usually represented as Q.
Specifically,
\[
g_{ab}\;(0,\lambda)\;=\;\exp{[\,A\;]}\;\exp{\left[  \;\xi\varrho\left(
t\right)  \;\right]  }\;\chi_{f}\;\langle\;\psi_{\lambda}(\,0\,)\;|\;\;L_{ab}%
\,|\;\psi_{\lambda}(\,0\,)\;\rangle,
\]%
\begin{equation}
A\,(\,t\,)\;\;=\;\;-\,\left(  \frac{m}{\left\vert \gamma\right\vert }\right)
^{2}t^{2}\,+\,c_{1}\,t\,+\,c_{2}\;\;,\qquad(c_{1},\,c_{2})\;\in\;C \label{A}%
\end{equation}

\medskip

where we have written the corresponding indices for the simplest supermetric
state solution,  $L_{ab}$ are the corresponding generators $\in Mp\left(
n\right)  $, and $\chi_{f}$ is coming from the odd generators of the big covering
group of symmetries of the specific model. 

\medskip

Considering for simplicity the
`square' solution for the three compactified dimensions  (spin
$\lambda$ fixed, $\xi\equiv-\left(  \overline{\xi}^{\overset{.}{\alpha}}%
-\xi^{\alpha}\right)  $), the exponential even fermionic part is given by:
\begin{align}
\varrho\left(  t\right)  \;\equiv\;\overset{\circ}{\phi}_{\alpha}\;\left[
\;\left(  \alpha\,e^{i\omega t/2}\right.  \right.   &  \left.  \left.
+\right.  \right.  \left.  \left.  \beta e^{-i\omega t/2}\right)  -\left(
\sigma^{0}\right)  _{\overset{.}{\alpha}}^{\alpha}\left(  \alpha e^{i\omega
t/2}-\beta e^{-i\omega t/2}\right)  \,\right] \\
&  +\frac{2i}{\omega}\left[  \left(  \sigma^{0}\right)  _{\alpha}^{\overset
{.}{\ \beta}}\ \overline{Z}_{\overset{.}{\beta}}+\left(  \sigma^{0}\right)
_{\ \overset{.}{\alpha}}^{\alpha}\ Z_{\alpha}\,\right]  \label{even}%
\end{align}

$\overset{\circ}{\phi}_{\alpha}, \, Z_{\alpha},\,\overline{Z}_{\overset
{.}{\beta}} $ being constant spinors, and $\alpha$ and $\beta$ are $\mathbb{C}%
$-numbers (the constant $c_{1}\in\mathbb{C}$ in Eq. (\ref{A}) due to the obvious physical
reasons and the chirality restoration of the superfield solution). 

By consistency, (and as in the string case), two geometric-physical options are
related to the orientability of the superspace trajectory  :
$\alpha = \pm\beta$. We take without loss of generality $\alpha= + \beta$ \,
then, exactly, there are two possibilities:

\medskip

{\bf(i)} The compact case which is associated to the small mass
limit (or $\left\vert \gamma\right\vert >>1)$ :

\begin{equation}
\varrho\left(  t\right)  \;=\,\left(
\begin{array}
[c]{c}%
\overset{\circ}{\phi}_{\alpha}\cos\left(  \omega t/2\right)  \;+\;\frac
{2}{\omega}Z_{\alpha}\\
-\overset{\circ}{\overline{\phi}}_{\overset{\cdot}{\alpha}}\sin\left(  \omega
t/2\right)  \;-\;\frac{2}{\omega}\overline{Z}_{\overset{.}{\alpha}}%
\end{array}
\right)  \label{comp}%
\end{equation}

\medskip

{\bf(ii)} The non-compact case, which can be associated to the imaginary
frequency ($\,\omega\,\rightarrow\,i\,\omega\,$: generalized inverted
oscillators) case:%

\begin{equation}
\varrho\left(  \,t\,\right)  \;\; = \; \;\left(
\begin{array}
[c]{c}%
\overset{\circ}{\phi}\,\cosh\,\left(  \, \omega\, t\,/2\,\right)  \;+
\;\frac{2}{\omega}Z_{\alpha}\\
-\;\overset{\circ}{\overline{\phi}}_{\overset{\cdot}{\alpha}}\,\sinh\,\left(
\, \omega\, t/2\,\right)  \;- \;\frac{2}{\omega}\,\overline{Z}_{\overset
{.}{\alpha}}%
\end{array}
\right)  \label{noncomp}%
\end{equation}
\newline Obviously (in both cases), this solution represents a
\textit{Majorana fermion} where the $\mathbb{C}$ (or $hypercomplex$) symmetry
 (wherever the case) is inside the constant spinors.

\medskip

The spinorial even part of the superfield solution in the exponent becomes:%

\begin{equation}
\xi\varrho\left(  t\right)  \;=\;\theta^{\alpha}\;\left(  \overset{\circ}%
{\phi}_{\alpha}\cos\left(  \omega t/2\right)  \;+\;\frac{2}{\omega}Z_{\alpha
}\right)  \;-\;\overline{\theta}^{\overset{\cdot}{\alpha}}\left(
-\,\overset{\circ}{\overline{\phi}}_{\overset{\cdot}{\alpha}}\sin\left(
\omega t/2\right)  \;-\;\frac{2}{\omega}\,\overline{Z}_{\overset{.}{\alpha}%
}\right)  \label{67}%
\end{equation}
\newline We easily see that in the above expression there appear a type of
continuous oscillation between the chiral and antichiral part of the bispinor
$\varrho(t)$, or \textit{Zitterbewegung} as shown qualitatively in
Fig. 1 for suitable values of the group parameters.

\begin{center}
\parbox[b]{4.478in}{\includegraphics[
height=3.245600in,
width=5.265800in,
height=2.7691in,
width=4.478in
]%
{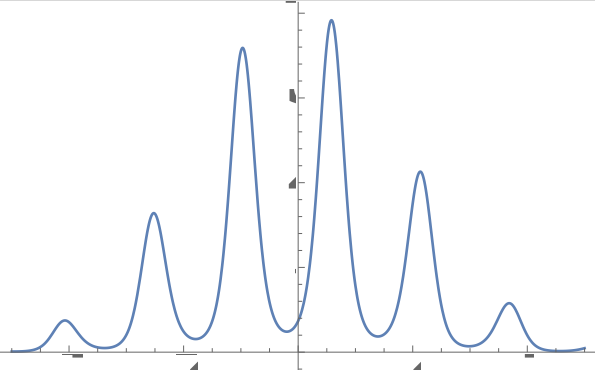}%
}

Figure 1: Chiral-antichiral oscillation (zitterbebegung) giving the pattern of
cat states from first principles. The asymmetry in the pattern can be seen,
marking a preferential temporal evolution.
\end{center}

\subsection{Generation and entanglement of Schrodinger cat states}

Let us remember that in general a "cat state" refers to a symmetric and
antisymmetric combination of coherent Heisenberg-Weyl
(HW) states with the
property that they sweep even and odd states of the harmonic oscillations. The
example of the definition in Ref \cite{dod} in terms of Heinserberg-Weyl displacement operators
are%
\[
\mathcal{D}_{\pm}\left(  \alpha\right)  \left\vert \;0\;\right\rangle \;
\equiv\;\left\vert \;\pm\;\alpha\;\right\rangle\;
\]%
\[
\frac{1}{2}\;\left[ \;\mathcal{D}\left(  \alpha\right) \; \pm\;\mathcal{D}\left(
-\alpha\right)  \;\right]  \;= \;\mathcal{D}_{\pm}\left(  \alpha\right)
\]
with the displacement operator standard definition
\[
\mathcal{D}\left(  \alpha\right)  \;=\;e^{\,\left(\,  \alpha\, a^{+}\, -\,\alpha^{\ast
}a\,\right)  }%
\]
We will now demonstrate, by comparing with the case of Ref.\cite{yurke},
that the theoretical construction and physical interpretation presented here
and in our previous works \cite{cirilo-sanchez},\cite{universe} \cite{NSPRD2021, NSPRD2023, Sanchez2019a,Sanchez2019b, Sanchez2019c}  is relevant from the fundamental point of view
as far as the very {\bf quantum structure of space-time} is concerned. 

\medskip

The starting
state in Ref. \cite{yurke} is a cat coherent state as described above, but in
our case here it is time dependent, the evolution operator being a Kerr type (non-linear) 
Hamiltonian  described in the expression Eq. (\ref{kerr}). To simplify the
problem, the authors of Ref. \cite{yurke} take a fixed interaction time 
$t = \pi/2\Omega$ where $\Omega$ is the coefficient of the anharmonic term:
\[
\left\vert\; \alpha,\ \pi/2\Omega\;\right\rangle\; =\;\frac{1}{\sqrt{2}}\;\left[\;
e^{-i\pi/4}\,\left\vert \;\beta\;\right\rangle \;+\;e^{i\pi/4}\;\left\vert \;-\beta
\;\right\rangle \;\right]
\]
with \,$\beta \,= \,\alpha \;e^{\,i\,\omega}.$

 Therefore, we observe that:

\medskip

{\bf (i)} The photoelectron or heterodyne number operator of Ref. \cite{yurke}
\ is simply%
\begin{equation}
\widehat{q}\;=\;\frac{1}{2}\;\left[\; \cos\,\left(\;  \theta\;+\;\omega \,t\;\right)  \left(
a+a^{+}\right)\;+\;i\,\sin\,\left(  \theta\;+\;\omega \,t\;\right)  \left( \, a\;-\;a^{+}\;\right)
\;\right] \; +\; q_{\,0} \label{q}%
\end{equation}
Let us compare with our operator in scalar form
\begin{align*}
\xi\,\varrho\left(  t\right)   &  \;=\;\xi\,\varrho\left(  t\right)  \;+\;\overline{\xi\,
}\overline{\varrho}\,\left(  t\right) \\
& \; = \;\cos\left(  \omega t/2\right)  \left(  \theta^{\alpha}\overset{\circ}%
{\phi}_{\alpha}+\overline{\theta}_{\overset{\cdot}{\alpha}}\overline{\phi
}^{\overset{\cdot}{\alpha}}\right)  +\sin\left(  \omega t/2\right)  \left(
\theta^{\alpha}\overset{\circ}{\phi}_{\alpha}-\overline{\theta}_{\overset
{\cdot}{\alpha}}\overline{\phi}^{\overset{\cdot}{\alpha}}\right)  +\;\frac
{4}{\omega}\left(  \theta^{\alpha}Z_{\alpha}-\overline{\theta}_{\overset
{\cdot}{\alpha}}\,\overline{Z}^{\overset{\cdot}{\alpha}}\right)
\end{align*}

Therefore, the identification is immediate between the solution field operators and
the photonic creation and annihilation operators in the case of the cited Ref. \cite{yurke}.

\bigskip

{\bf (ii)} The nonlinearity of the Kerr term (nonlinear optical medium) introduces
the ingredient $SU(1,1)$ and the covering $Mp(2)$ in the simplest case, namely
(remember that $\Omega$ is the oscillator non-linearity parameter and takes a
specific form when the physical setting is defined)
\;
\begin{equation}
H \; = \;\omega\;\widehat{n} \;+ \;\Omega\;\widehat{n}^{2} \;= \;\left(\; \omega \; + \;\Omega \;\right)
a^{+}a \;+\; \Omega\; a^{+2}a^{2} \label{kerr}%
\end{equation}

The evolution generator in our case contains all the elements of the group
$Osp(n)$ (or its covering $OMp(n)$), which is the largest symmetry harmonic
oscillator as far as the $a$  variables are concerned, with the odd part
given by the generators: $F_{+}=\frac{1}{2}a^{+}$ and $F_{-} = \frac{1}{2}a$

\begin{center}
\raisebox{-0pt}
{\includegraphics[
height=5.701700in,
width=6.096900in,
height=4.2428in,
width=4.5359in
]%
{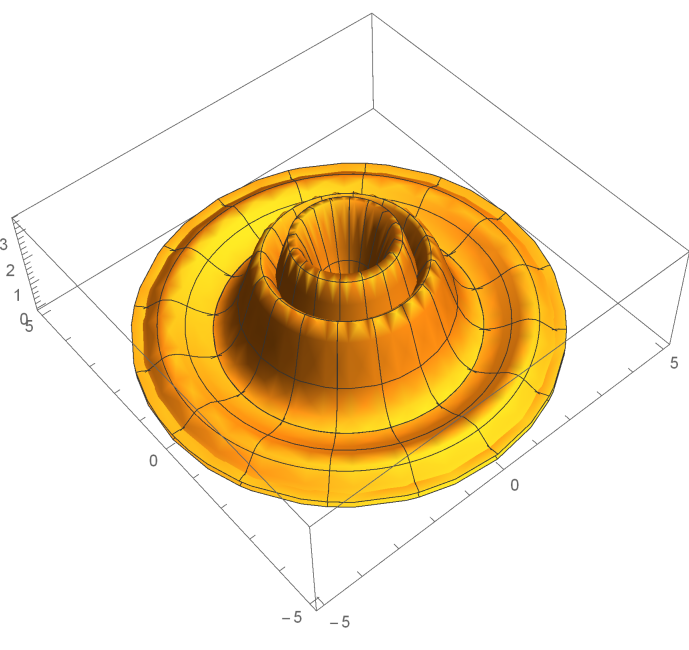}%
}

Figure 2: \ 3\,D picture of the chiral-antichiral oscillation (cat states
pattern)
\end{center}

\subsection{Entanglement of coherent states and evolution of probability}

We consider SU(1,1) states as an example, namely%
\begin{align*}
\left\vert \;k,\zeta\;\right\rangle  &  \;=\;e^{\;(\,\xi\, K_{+}-\xi^{\ast}K_{-})}\;\left\vert \;
k,\;0 \;\right\rangle \\
&  = \;\left( \; 1\;- \;\left\vert \; \zeta\;\right\vert ^{2}\;\right)  ^{k}e^{\;\zeta \,K_{+}%
}\;\left\vert \;k,\;0\;\right\rangle
\end{align*}
Over the orthonormal basis $\left\vert k,m\right\rangle $ we have%
\[
\left\vert \;k,\zeta\;\right\rangle \;=\;\left(\;  1\;-\;\left\vert \;\zeta\;\right\vert
^{2}\;\right)^{k}\underset{m=0}{\overset{\infty}{\sum}}\;\left[\;  \frac
{\Gamma\left( \; m\;+\;2k\;\right)  }{m!\;\Gamma\left( \; 2k\;\right)  }\;\right]  \zeta
^{m}\;\left\vert \;k,\,m\;\right\rangle
\]
where 
$$\xi\;= \;-\;\left(\,\chi/2\,\right)\,e^{-i\varphi},\, \;\;\; \zeta \,=\,\left(\,
\xi/\,\left\vert \,\xi\,\right\vert \,\right) \,tanh\left\vert \,\xi \,\right\vert
\;= \;-\,tanh \,\left( \, \chi/2\,\right)  \,e^{-\varphi}$$ 

Consequently, the parameter $\zeta$
is restricted to $\left\vert \,\zeta\,\right\vert <1$ (disk) and $(\chi,\varphi)$ is
coming from the parametrization of the quotient space is $SU(1,1)/U(1)$ (the
upper sheet of the two-sheet hyperboloid), and the standard coherent state is
specified by a unit pseudo Euclidean vector: \
\[
\widehat{n}\;=\;\left(  \sinh\chi\cos\varphi,\;\;\sinh\chi\sin\varphi,\;\;\cosh
\chi\right)
\]
The overlap
\[
\left\langle \;k,\;\zeta_{a}\;\right.  \left\vert \;k,\;\zeta_{b}\;\right\rangle \;=\;\left(\;
1 - \;\left\vert \;\zeta_{a}\;\right\vert ^{2}\;\right)^{k}\;\left(\;  1 \;-\;\left\vert\;
\zeta_{b}\;\right\vert ^{2}\;\right)  ^{k}\left(\;  1\;-\;\zeta_{a}^{\ast}\;\zeta
_{b}\;\right)^{-2k}
\]

\bigskip

Then, an orthonormal state \cite{semi} to 
$\left\vert \;k,\zeta_{\,b}\;\right\rangle$
  is:

\[
\left\vert \,k,\,\widetilde{\zeta_{\,b}}\,\right\rangle \;\equiv \;\left(  1-\left\vert\,
\zeta_{b}\,\right\vert ^{2}\right)  ^{k}\;\underset{m=0}{\overset{\infty}{\sum}%
} \;\left[\;  \frac{\Gamma\left(  m+2k\right)  }{m\,!\;\,\Gamma\left(  2k\right)
}\;\right]  \left[\,  \zeta_{b}^{m}+\frac{\left(  1-\left\vert \,\zeta
_{a}\,\right\vert ^{2}\right)^{k}}{\left(  1-\zeta_{a}^{\ast}\;\zeta_{b}\right)
^{2k}}\;\,\zeta_{a}^{m}\;\right]  \left\vert \,k,m\,\right\rangle
\]

\bigskip

where to simplify we denote as semi-coherent state the following:%

\[
\left\vert \;\widetilde{\alpha}\;\right\rangle \;=\;\frac{\left\vert \;\alpha\;
\right\rangle -\left\vert \;\beta\;\right\rangle \left\langle \;\beta\right.
\left\vert\; \alpha\;\right\rangle }{\sqrt{1\;-\;\left\vert \left\langle \;\beta\;\right.
\left\vert\; \alpha\;\right\rangle \right\vert ^{2}}}%
\]

 \bigskip

To make a construction like Bell's for coherent states one orthonormalizes the
SU(1,1) states for example to have a basis of the standard type $\left\vert
1\right\rangle $ and $\left\vert 0\right\rangle $, namely $\left\vert
\alpha\right\rangle ,\left\vert \widetilde{\alpha}\right\rangle $ (with $k$
fixed) consequently considering $\left\vert \alpha\right\rangle $ 
$\left\vert \widetilde{\alpha}\right\rangle $,   ($\alpha$, $\widetilde{\alpha}$ denoting the coherent state eigenvalues). \ They can be used for computing
processes because they are orthonormal:

\[
\left\vert\; \psi\;\right\rangle \; = \;\frac{\left\vert \;\alpha\;\right\rangle
\;\otimes \;\left\vert \; \widetilde{\alpha}\;\right\rangle \;\; +\; \; e^{i\,\varphi}\left\vert\;
\widetilde{\alpha}\;\right\rangle \;\otimes \;\left\vert \;\alpha\;\right\rangle }%
{\sqrt{\;2\;}}%
\]

\bigskip

(Let us notice from the expressions of the coherent state as a function
of $\left\vert k,m\right\rangle $ that the Bargmann index must be the same). 

\medskip

The Figures 3 below show the dynamics of the probability for $\left\vert \widetilde
{\alpha}\right\rangle $ which is a gaussian-type \cite{mult} entangled state
where the degree of entanglement varies as a function of time.

\bigskip

\begin{center}
\raisebox{-0pt}{\includegraphics[
height=4.359500in,
width=7.301600in,
height=2.5157in,
width=4.2013in
]%
{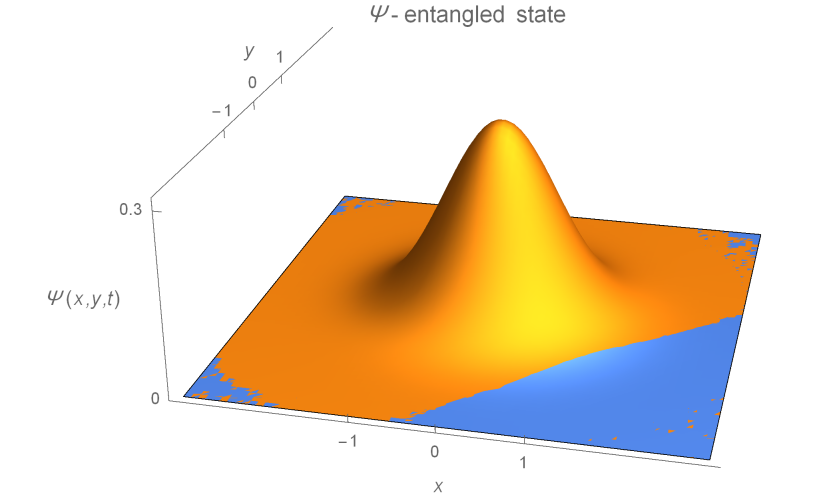}%
}

\raisebox{-0pt}{\includegraphics[
height=4.386300in,
width=7.336200in,
height=2.5711in,
width=4.2895in
]%
{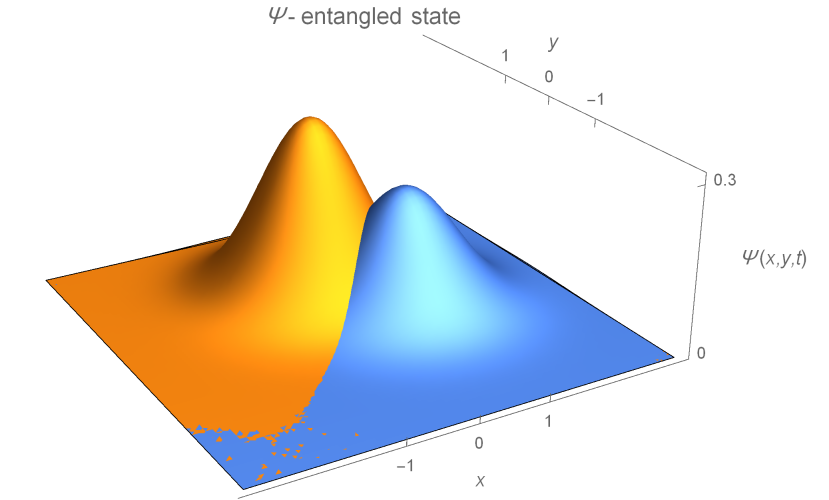}%
}

\raisebox{-0pt}{\includegraphics[
height=4.350900in,
width=7.282600in,
height=2.5486in,
width=4.2566in
]%
{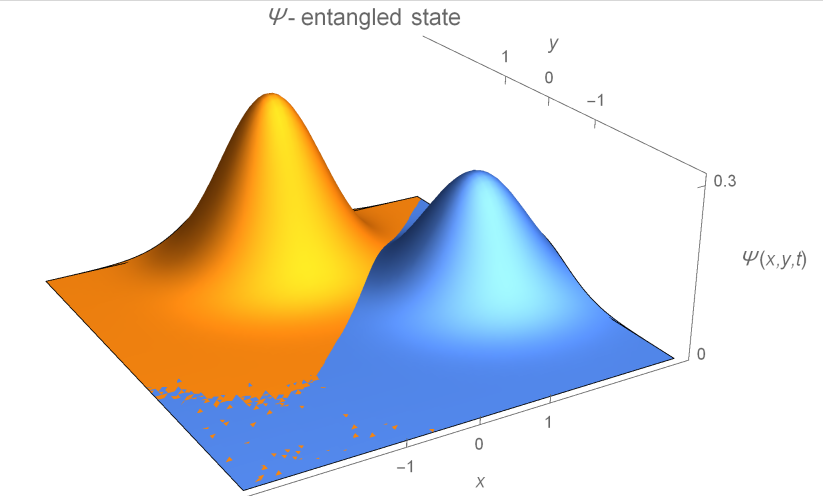}%
}

Figures 3: The three images show from top to down the entangled coherent state where the degree of
entaglement varies as a function of time from the highest to the lowest degree controlled by
the overlap $\left\langle\alpha \, \vert \,\beta \right\rangle$. Here
$(x,y)$ correspond to $\operatorname{Re}\,\widetilde{\alpha}$ and \,$ \operatorname{Im}\,\widetilde{\alpha}$.
\end{center}

\bigskip

By considering that the evolution equation is of Fokker-Planck type due to the fact that 
the Hamiltonian is of the type of Eq. (\ref{kerr}),  we have:

\[
\frac{\partial \,P}{\partial\, t}\;\;=\;\;\frac{\epsilon}{2}\;\;\underset{i}{\sum}\;\left[\;\,
\frac{\partial\left(  w_{i}\,P\right)  }{\partial\, w_{i}}\;\;+\;\;\frac{1}{2}%
\;\frac{\partial^{2}\left(  w_{i}\,P\right)  }{\partial\, w_{i}^{2}}\;\,\right]
\]

\medskip

with $$w_{i}\; = \;\operatorname{Re}\,\alpha, \;\; \operatorname{Im}\,\alpha,\;\;\;\operatorname{Re}\,
\widetilde{\alpha},\;\;\;\operatorname{Im}\,\widetilde{\alpha},\;\;\;\left( \; i = 1,..,4\;\right)
, \; \; \;\widetilde{\alpha}\; = \;\;\frac{\alpha-\beta\,\left\langle \,\beta\right.
\left\vert \,\alpha\,\right\rangle }{\sqrt{1\,-\,\left\vert\, \left\langle \,\beta\right.
\left\vert \,\alpha\,\right\rangle \,\right\vert ^{2}}},$$ 
and in standard form the
probability:
$$P\left( \; \alpha,\;\alpha^{\ast},\;\widetilde{\alpha},\;\widetilde
{\alpha}^{\ast}\,,\,t\;\right)  \;\;=\;\;\left\vert \;\left\langle \; m,\,n\;\right.  \left\vert\;
\psi,\,k\;\right\rangle \;\right\vert ^{2}.$$

\section{Schrodinger Cat States and Mp(2): Even and Odd Sectors }

From the point of view of the considered Hilbert space, divided into even and
odd states we have the following apparent correspondence between standard cat
states and the basic states of Mp(2), namely $\left\vert 1/4\right\rangle $
and $\left\vert 3/4\right\rangle :$

\begin{equation}%
\begin{tabular}
[c]{|l|l|l|l|}\hline
$Heisenberg-Weyl$ \ \ \ \ \ \ \ \ \ \ \ \ \ \ \ \ \ \  &  & $Metaplectic$
$Mp(2)$ \ \ \ \ \ \ \ \ \ \ \ \ \ \ \ \ \ \ \ \ \  & \\\hline
$\ \ \ \ \ \ \ \ \ \ \ \ \ \ \ \ \ \left\vert \alpha_{+}\right\rangle $ &
$\longrightarrow$ & $\ \ \ \ \ \ \ \ \ \ \ \ \left\vert 1/4\right\rangle $ &
Even\\\hline
$\ \ \ \ \ \ \ \ \ \ \ \ \ \ \ \ \ \left\vert \alpha_{-}\right\rangle $ &
$\longrightarrow$ & $\ \ \ \ \ \ \ \ \ \ \ \ \left\vert 3/4\right\rangle $ &
Odd\\\hline
\end{tabular}
\ \ \ \ \label{t}%
\end{equation}

\bigskip

where explicitly the standard cat Schrodinger states are%

\[
\left\vert \alpha_{\pm}\right\rangle \;=\;\frac{1}{\sqrt{2\;\pm\;2\;e^{-2\left\vert
\alpha\right\vert ^{2}}}}\;\left[ \; \left\vert \;\alpha \;\right\rangle \;\pm\;\left\vert\,
-\alpha\,\right\rangle \,\right]
\;\]
\\
The situation from the point of view of the density matrices is clear in favor
of the $Mp(2)$ states that follow the principle of minimum representation. In
the case of the standard cat states, the density matrices
for the even and odd sectors are the following

\begin{equation}
\rho_{\pm\alpha}\;=\;\frac{1}{2\;\left(\;  1\;\pm \;e^{\;-2\;\left\vert \;\alpha\;\right\vert\;
^{2}}\;\right)  }\;\left[\; \left\vert\, \alpha\,\right\rangle \left\langle\,
\alpha\,\right\vert \,+\,\left\vert\, -\,\alpha\,\right\rangle \left\langle \,-\,\alpha\,
\right\vert \,\pm \,\left(\,  \left\vert\, -\,\alpha\,\right\rangle \left\langle
\,\alpha\,\right\vert \,+\,\left\vert \,\alpha\,\right\rangle \left\langle \,-\,\alpha \,
\right\vert \,\right) \, \right]  \label{+-}%
\end{equation}
\\
but in the fundamental case of the minimal group representation Mp(2), we evidently have
a diagonal representation (as described by Sudarshan) clearly differentiated
into the corresponding {\it even} and {\it odd} subspaces : %
\begin{equation}
\rho_{Mp\left(  2\right)  }\; = \;\left\{
\begin{array}
[c]{c}%
\left\vert \;1/4\;\right\rangle \left\langle \;1/4\;\right\vert \;\;\rightarrow \; \;even\\
\left\vert \;3/4\;\right\rangle \left\langle \;3/4\;\,\right\vert \;\;\rightarrow \; \;odd
\end{array}
\right.  \label{rmp}%
\end{equation}
\\
 Looking at equation (59) and comparing it with equation (58) for the even or odd cases
we see that the fundamental substrate of space time following the principle of
minimum group representation is described by the Metaplectic group
and not by the given standard Schrodinger cat states: this situation is
clearly seen from the tail in the expression for the $\rho_{\pm\alpha}$ Eq. (58) while such tail does not appear in the fundamental expressions of $\rho_{Mp\left( 2\right)}$ for both $Mp\left( 2\right)$ even and odd states.

Consequently, in the case of this last description, Eq.(\ref{rmp}), we see that the geometric substrate of spacetime is {\it quantum in
essence} because we are in the minimal representation and it is
diagonal (classical and quantum aspects are both represented), and  where  the 
power of coherent states in describing {\it both continuum and discrete
space-time} is completely manifest.
It must be also  noticed:

\medskip

{\bf(i)} \; The states $\left\vert 1/4\right\rangle ,\left\vert 3/4\right\rangle$  are not
the same standard cat states as from Ref. \cite{yurke} but are a fundamental part of
the very structure of the spacetime itself and do not require an extrinsic
generation process.

 \medskip
 
{\bf(ii)}  The basic states could be forming a
generalized state of the type

\[
\left\vert \Psi_{Mp(2)}\right\rangle _{gen}\;=\;\frac{1}{\sqrt{\left\vert
A\right\vert ^{2}\;\pm \;\left\vert B\;\right\vert ^{2}}}\;\left[\;  A\;\left\vert\,
1/4\,\right\rangle \;\pm \;B\;\left\vert \,3/4\;\right\rangle \;\right]
\]
\\
(similar to the standard Schrodinger cat state $\rho_{\pm\alpha}$) giving in
this case the following density matrix:
\begin{equation}
\rho_{Mp(2)_{gen}}=\frac{1}{\left\vert A\right\vert ^{2}\pm\left\vert
B\right\vert ^{2}}\left[\,  \left\vert A\right\vert ^{2}\;\left\vert
1/4\right\rangle \left\langle 1/4\right\vert \,+\,\left\vert B\right\vert
^{2}\;\left\vert 3/4\right\rangle \left\langle 3/4\right\vert 
\,\pm\,\left(
B^{\ast}A\,\left\vert 3/4\right\rangle \left\langle 1/4\right\vert +A^{\ast
}B\left\vert 1/4\right\rangle \left\langle 3/4\right\vert\right)\,\right]
\end{equation}

This expression is comparable in shape to the density matrix $\rho_{\pm\alpha}$ where the parameters $A$ and $B$ may be
subject to extrinsic control conditions determined by Majorana Dirac
type equations as we established in references 
\cite{universe}, \cite{cirilo-sanchez}.

\section{Entanglement in Quantum de Sitter Space-Time and Black Hole Space-Times}

In this Section we implement the entanglement results obtained in this paper in two important quantum gravitational examples: de Sitter and Black Hole quantum space-times.
The basis of these quantum space-time descriptions can be found in Refs \cite{NSPRD2021}, \cite{NSPRD2023}, 
\cite{Sanchez2019a} 
\cite{Sanchez2019c},\cite {cirilo-sanchez} 

\bigskip

The {\bf de Sitter} space-time admits at both the classical and quantum levels a \textit{inverted}, (ie with imaginary
frequency) harmonic oscillator description, where the \textit{oscillator constant $\kappa_{osc} \; =  \;c/l_{osc}$} and {oscillator length} $ l_{osc}$
are given by \cite{NSPRD2021},\cite{Sanchez2019c}: 

\begin{equation}  \label{l-deS}
l_{osc\;dS}^{-2} \;= \; \left(\,\frac{ m \omega}{\hbar}\,\right)_{dS} \; =
\; H^2 \;=  \; \frac{\,8\,\pi\,G\, \Lambda}{3}
\end{equation}

\bigskip

The \textit{oscillator length} $l_{osc}$ is classically the Hubble radius, the Hubble
constant $H = \kappa$ being the \text{surface~gravity}, as the black hole
surface gravity is the inverse of (twice) the black hole radius.

\bigskip

For \textbf{Anti-de Sitter} space-time, the description is similar and, (because the AdS two sheet hyperbolid embedding in Minkowski space-time with respect to the deS one sheet hyperboloid),  Anti-de Sitter is associated to the real frequency \textbf{(non inverted)} harmonic oscillator.

\bigskip

For the (Schwarzschild) \textbf{black hole} space-time description, the physical
magnitudes as the oscillator constant and the oscillator length are related
to the black hole mass $M$ by: 

\begin{equation}\label{l-BH}
l_{osc\; BH}^{-2} \;= \; \left(\,\frac{ m \omega}{\hbar}\,\right)_{BH} \; =
\;\, l_P^{-2} \,\left(\,\frac {m_P}{M} \,\right)^2
\end{equation}

\bigskip

$l_P$ being the Planck length and $m_P$ the Planck mass:
\begin{equation*}
l_P \;= \; (\,2 \,G\,\hbar\, /\,c^3 \,)^{1/2}, \, \qquad h_P \;= \; c\,/\,l_P
\end{equation*}

 \bigskip

The {\bf discrete} states and their spectrum  describe the quantum space-time levels.

Interestingly, the Metaplectic group states with its {\it both} sectors and discrete representations, $\vert\, 2\,n \,\rangle$ \,and \, $\vert \,2\,n+1 \,\rangle$, {\it even} and  {\it odd} states, fully {\it cover} the {\it complete} Hilbert space $\mathcal{H}$:  
\begin{equation}
\mathcal{H}\;\; =\;\; \mathcal{H}_{\,(\, + \,)} \;\; \oplus\;\;\mathcal{H}_{\,(\,- \,)} 
\end{equation}

The $(\,\pm \,)$ symmetric and antisymmetric sum of the two kind ({\it even} and {\it odd})  
states provides the {\it complete} covering of the Hilbert space and of the space-time mapped from it:  
\begin{equation}
\Psi \left(n\right)  \;\; = \;\; \Psi^{\left(+ \right)} \left(
2n\right) \;\; + \;\;\Psi^{ \left(- \right)} \left( 2n + 1\right) 
\end{equation}

The complete covering of the Hilbert space with the complete covering of the quantum space-time is realized by the Metaplectic group symmetry which equivalently provides the CPT symmetric states and unitarity states. For a different approach in the searching of these properties in black hole states see eg Refs \cite{tHooft1}, \cite{tHooft2022}.  

\bigskip

The Classical-Quantum Duality of the space-time is also realized in the $ Mp(n)$ symmetry, because the complete covering, the global complete space-time (and full phase space completion) is needed to make manifest the classical-quantum duality of the space-time (and its phase space mapped from it) .

\begin{itemize}

\item {It is worth to mention that similar discrete levels can be obtained from the 
 global (complete) classical - quantum  duality including gravity \cite{Sanchez2019a}, \cite{Sanchez2019b} , \cite{NSPRD2021}, namely classical-quantum gravity duality:   The two {\it even} and {\it odd}  (local) carts or sectors and their (global) $(\pm)$ sum of states, reflect  a relation between the $ Mp(n)$ symmetry and the classical-quantum duality.} 

\item {The two $\sqrt{\,(2n+1)}$ and  $\sqrt{\,2n}$, {\it even} and {\it odd} sets are local coverings and they are {\it entangled} one to each other. The symmetric or antisymmetric sum of these sectors are {\it global} and they are required to cover the {\it whole} manifold.}  

\item {Moreover, the corresponding $(\,\pm\,)$ global states are complete, CPT symmetric and unitary, the levels $n = 0, 1, 2, ....,$ cover the whole Hilbert space $\mathcal{H}\; =\;\mathcal{H}_{\,(\, + \,)} \;\; \oplus\;\;\mathcal{H}_{\,(\,- \,)}$ and all the space-time regimes.}

\item {The total $n$ states range over \textit{all} scales from
the lowest excited levels to the highest excited ones covering the two dual branches $(+)$ and $(-)$ or Hilbert space sectors and corresponding space-time coverings.
The two $(+)$ and $(-)$ dual sectors are entangled.}

\end{itemize}

This is interesting because the classical-quantum gravity duality allows that signals or states in the quantum gravity (trans-Planckian) domaine, or semiquantum gravity (inflationary) domaine, do appear as low energy effects in the semiclassical/classical universe today. 

From the results of this paper, we have seen that such gravity or cosmological domains {\it are entangled }: The quantum trans-Planckian primordial phase is a quantum constant curvature de Sitter phase, followed by a quasi-de Sitter (inflationary) phase, and the late Universe today is a  semiclassical and classical gravity de Sitter phase. The most quantum 
 and primordial (trans-Planckian) period and its dual: the most classical and late (today) one are {\it entangled}.

\medskip

 As is well known, de Sitter space-time is the realistic phase of the accelerated expansion of our universe today as shown from the robust set of observational data,  cosmo and astro observations. de Sitter or quasi-de Sitter space-time is also the cosmic inflationary phase for the early universe.  In between these two asymptotic stages, the radiation and matter dominated cosmological eras are described by the well known non accelerated expansions, eg classical or semiclassical (non quantum) space-time. Entanglement between states of several kinds (eg radiation fields, matter fields,  semiclassical geometry) can occur in such stages, and with different degrees of intensity and variability.  Enhancement and conditions of the entanglement in such stages depend too of other effects (such as the presence of additional thermal features, magnetic fields or other fields and sources). 

\medskip

 Evidently, the stability of the entanglement will depend on the coherence of
the states that form it. Consequently, if the symmetries in the generation of the coherent states are preserved, the entanglement will survive. The entanglement between the asymptotic in and out de Sitter stages of the universe are well suited examples of it, de Sitter or quasi de Sitter stages being maximally symmetric space-times,  non singular and of constant curvature.

\medskip

 Semi-classical (or semi-quantum) de Sitter or quasi-de Sitter space-time is the cosmic inflationary phase for the early universe in the modern Standard Model of the Universe that is supported on theoretical grounds by the General theory of Relativity for gravity and QFT (Quantum Field Theory) for matter and particles, and on observational grounds by a wide set of cosmic and astro concordant observations:  CMB (Cosmic Microwave Background), LSS (Large Scale Structure) among other data.

\medskip

 Quantum de Sitter space time for the early universe does appear as the earlier phase from which cosmic de Sitter inflation is a continuation, it does also appear on the grounds of the classical-quantum gravity duality, its connection to the generalized (quantum) Synge algebra, and in the description of quantum space-time in terms of the generalized coherent states of the Metaplectic group.

 \section{Remarks and Conclusions}

The results of this paper have both:  fundamental and practical implications of quantum physics {\it from a novel perspective}:

 \bigskip
 
{\bf(i)} For the quantum space-time structure properties from one side, by studying in its description concepts as the Berry phase and more generally geometrical (Berry type) phases which untill now have been most purely studied in quantum systems but not in quantum gravity as we do here: This is in the quantum  space-time context including its trans-Planckian domain and the role of the {\it Metaplectic group} which is non compact in this case. 

\bigskip

{\bf(ii)} Second, the new results on entanglement with the states of the Metaplectic group and its covering, which can be taken into account for the searching of new measurable signals: 
being them from black holes, the gravitational wave domaine and the high energy domain, or the de Sitter primordial phases (inflation and before inflation),  and the late de Sitter cosmological vacuum  (today dark energy).

\bigskip

{\bf(iii)}  Some points here considering the coherent state solutions and the entangled case are the following: 
 As we have been pointed out in Ref \cite {PLB2008}, a remarkable property of the simple solution given by the physical state $g_{ab}(x)$ is that it is localized in a particular position of the space–time. The supermetric coefficients $a$ and $a*$ play the important role of localize the fields in the bosonic part of the superspace,  in a similar and suggestive form as the well-known “warp factors” of multidimensional gravity for a positive (or negative) tension brane. 
 
 But the essential difference here is due to the c-numbers $a$ and $a*$  coming from the $B_0$ (even) fermionic part of the superspace under consideration. Therefore, not additional and/or topological structures that break the symmetries of the model (i.e., the reflection $Z_2$-symmetry) are required in our description: The natural structure of the superspace does produce this effect due to the symmetries of the Metaplectic group. 

\bigskip

{\bf(iv)} The Coherent  (Gaussian type) solution  is a very well-defined physical state in any Hilbert space  from the mathematical point of view, contrarily to the case in the literature, e.g \,$u(y) = c \; \exp\, (-\,H\,|\,y\,|\,)$  \cite{bajc2000} and references therein. In  such a case,  it was possible to find a manner to include it in any Hilbert space, but it was strongly needed to take special mathematical and physical particular assumptions whose meaning are not clear.

\bigskip

{\bf(v)} In the entanglement case, as we can see in the Figures (Fig. 2 and Figs 3), all the above properties are preserved. The locality is subject to the degree of entanglement and the limit when $a\rightarrow \infty$ where the Gaussian condition (envelope) is lost. In such a limit, only the odd (fermionic) part of the super-manifold survives.

\bigskip

{\bf(vi)} There exists a clear evidence of a \textit{time arrow } coming from the physical states as a bilinear combination of the basic states of the Metaplectic $Mp(n)$ group. This appears in the  appreciable asymmetry displayed by Figure (1) "Zitterbebegung" as in the fact (CPT) that: 
\[
\left\langle \;\Psi_{3/4}\left(  t\right)  \;\right\vert\; L_{ab}\;\left\vert\;
\Psi_{1/4}\;\left(  t\right)  \;\right\rangle \;\neq \;\left\langle \;\Psi_{1/4}\left(
t\right) \; \right\vert \;L_{ab}\;\left\vert \;\Psi_{3/4}\left(  t\right)\;\right\rangle
\]
 
 \medskip
 
{\bf(vii)} The obtained Berry phase applied to the de Sitter inflation case is {\it imaginary} describing the inflationary exponential factor acceleration, as it must be.
 We also consider the case of cosmic perturbations in the slow roll regime and related  the Berry phase to the cosmological observables: scalar and tensor spectral indices $ n_s$ and $n_T$ and the ratio of tensor to scalar perturbations) .  

 \bigskip

{\bf(viii)}  From the density matrix viewpoint in the entanglement context, the precise relation between the Schmidt type representation and the physical state fulfilling the Minimal Group Representation Principle (MGRP), that is bilinear in the basic states of the $Mp(n)$ group, is found. 
 The mapping for the physical state refers to a {\it new non-diagonal} coherent state representation complementary to that of the Sudarshan diagonal representation. 
 
 \bigskip
 
 {\bf(ix)}  In the number basis $| n >$ the physical state corresponding to $s = 2$ (graviton, related to the space-time metric) provides the discretization of the space- time for small $n$, going to the continuum for $n \rightarrow \infty$. 
 
 \bigskip
 
 {\bf(x)} The basic states in the Minimal Group Representation  sense: $|1/4>$ and $|3/4>$ (belonging to the even and odd sectors of the Hilbert space respectively) are a fundamental part of the very structure of the space-time itself and do not require an additional extrinsic generation process as in the standard Schrodinger cat states and their entanglement. 
 
 \bigskip

{\bf(xi)} The entanglement results in quantum de Sitter space-time admit at both the classical and quantum levels an inverted, (ie. with imaginary frequency) harmonic oscillator description, (a real frequency (non inverted) harmonic oscillator in AdS). 
 
\bigskip

{\bf(xii)}  In the entanglement results for the (Schwarzschild) black hole space-time, 
the physical magnitudes as the oscillator constant and the oscillator length are related to the black hole mass $M$ and the Planck mass $m_P$ :
 $l_{osc\; BH}^{-2} \;= \; \left(\, m \omega /\hbar\,\right)_{BH} \; =
\;\, l_P^{-2} \,\left(\, {m_P}/{M} \,\right)^2$.

\bigskip

The external and internal  regions of the black hole are classical-quantum duals of each other and are {\it entangled}. The entanglement occurs from the continuum external semiclassical/classical states and the discrete very quantum, Planckian and trans-Planckian states.  Discrete states here describe the quantum space-time levels and their spectrum. 

\bigskip

{\bf(xiii)}  Gravity or cosmological domains from one side and the other of the Planck scale are entangled: The quantum trans-Planckian primordial de Sitter phase (followed by a quasi-de Sitter (inflationary) phase), and the late Universe today semiclassical and classical gravity de Sitter phase are  dual of each other and {\it are entangled}.
This is interesting because the classical-quantum gravity duality allows that signals or states in the quantum gravity (trans-Planckian) primordial domaine do appear as low energy effects in the semiclassical/classical gravity universe today. 
Such effects deserve future investigations.

\bigskip

\section{Acknowledgements}

\bigskip

DJCL acknowledges the institutions CONICET and the Keldysh Institute of Applied Mathematics and the support of the Moscow Center of Fundamental and Applied Mathematics,Agreement with the Ministry of Science and Higher Education of the Russian Federation, No. 075-15-2022-283

\end{document}